\documentstyle[aps,multicol,pra,eqsecnum,epsfig]{revtex}
\begin{document}
\draft
\title{Quantum jumps in a two-level atom}
\author{H.M. Wiseman$^{1}$\footnote{Electronic address: 
wiseman@physics.uq.edu.au} and G.E. Toombes$^{1,2}$}
\address{$^{1}$Centre for Laser Science, 
Department of Physics, The University of Queensland, Queensland 4072 Australia\\
$^{2}$Department of Physics, Cornell University, Ithaca, New York 14853-2801 U.S.A.
}
%\date{July 3, 1998}
\maketitle

\begin{abstract}
	A strongly-driven ($\Omega \gg \gamma$) 
	two level atom relaxes towards an equilibrium state 
	$\rho$ which is almost completely mixed. One interpretation of this state is 
	that it represents an ensemble average, and that an individual atom 
	is at any time in one of the eigenstates of $\rho$. The theory of 
	Teich and Mahler [Phys. Rev. A {\bf 45}, 3300 (1992)] makes this 
	interpretation concrete, with an individual atom jumping 
	stochastically between the two eigenstates when a photon is emitted. 
	The dressed atom theory is also supposed to describe the quantum 
	jumps of an individual atom due to photo-emissions. But the two 
	pictures are contradictory because the dressed 
	states of the atom are almost orthogonal to the eigenstates of $\rho$. In 
	this paper we investigate three ways of measuring the field radiated 
	by the atom, which attempt to reproduce the simple quantum jump 
	dynamics of the dressed state or Teich and Mahler models. 
	These are: spectral detection (using optical 
	filters), two-state jumps (using adaptive homodyne detection) and 
	orthogonal jumps (another adaptive homodyne scheme). 
	We find that the three schemes closely mimic the jumps of 
	the dressed state model, with errors of order $\frac34 
	(\gamma/\Omega)^{2/3}$, $\frac14 (\gamma/\Omega)^{2}$, and 
	$\frac34 (\gamma/\Omega)^{2}$ respectively. The significance of this 
	result to the program of environmentally-induced superselection is 
	discussed.
 
\end{abstract}

\begin{multicols}{2}

\newcommand{\beq}{\begin{equation}} 
\newcommand{\eeq}{\end{equation}}
\newcommand{\bqa}{\begin{eqnarray}} 
\newcommand{\eqa}{\end{eqnarray}}
\newcommand{\nn}{\nonumber} 
\newcommand{\nl}[1]{\nn \\ && {#1}\,}
\newcommand{\erf}[1]{Eq.~(\ref{#1})}
\newcommand{\dg}{^\dagger}
\newcommand{\rt}[1]{\sqrt{#1}\,}
\newcommand{\smallfrac}[2]{\mbox{$\frac{#1}{#2}$}}
\newcommand{\half}{\smallfrac{1}{2}} 
\newcommand{\bra}[1]{\langle{#1}|} 
\newcommand{\ket}[1]{|{#1}\rangle}
\newcommand{\ip}[1]{\langle{#1}\rangle}
\newcommand{\sch}{Schr\"odinger } 
\newcommand{\schs}{Schr\"odinger's }
\newcommand{\hei}{Heisenberg } 
\newcommand{\heis}{Heisenberg's }
\newcommand{\bl}{{\bigl(}}
\newcommand{\br}{{\bigr)}} 
\newcommand{\ito}{It\^o }
\newcommand{\str}{Stratonovich } 
\newcommand{\dbd}[1]{\frac{\partial}{\partial {#1}}}

\section{Introduction}

The quantum jump, the effectively instantaneous transition of an atom 
from one state to another, was the first form of nontrivial 
quantum {\em dynamics} to 
be postulated \cite{Boh13}. Of course Bohr's theory did not survive the quantum 
revolution of 
the 1920s. In particular, the idea of jumps 
appeared to be in sharp conflict 
with the continuity of \schs wave mechanics \cite{Hei67}. 
In the aftermath of the revolution, quantum jumps were revived 
\cite{Dir30} with a new interpretation as state reduction caused by 
measurement. But Wigner and Weisskopf \cite{WeiWig30} had already derived
 the exponential decay of spontaneous emission from the coupling 
 of the atomic dipole to 
the continuum of electromagnetic field modes. That is, 
they did not require the hypothesis of quantum jumps. 
Later, more sophisticated theoretical techniques, such as 
the master equation, were developed for dealing with the irreversible 
 dynamics of such open quantum systems \cite{Blo46,Zwa60,Lou73}. 
 In the master equation 
 description, the atom's state evolves smoothly and deterministically. 
 Perhaps as a consequence, interest in quantum jumps as a way of 
describing of atomic dynamics faded.

Quantum jump models for atoms were never entirely forgotten;  
the dressed state 
model \cite{CohRey79} was used successfully to give an 
intuitive explanation of the 
Mollow triplet \cite{Mol69} in resonance fluorescence. 
However, it was 
the electron shelving experiments of Itano and co-workers \cite{Ber86} 
which refocussed 
attention on the {\em conditional} dynamics of individual atoms.
Subsequent work on waiting time distributions \cite{CohDal86,ZolMarWal87}
led to a renewal of interest in quantum jump descriptions 
\cite{CarSinVyaRic89}. It was shown by Carmichael \cite{Car93b} that 
quantum jumps are an implicit part of standard photodetection theory. 
This link between continuous quantum measurement theory and
stochastic quantum evolution for the pure state of the system 
was considered by many other workers 
around the same time and subsequently 
\cite{GarParZol92,WisMil93a,WisMil93b,WisMil93c,Wis93a,Wis94a,Car93a,%
CarKocTia94,GarKni94,GoeGra94b}.
Independently, Dalibard, Castin and M\o{}lmer \cite{DalCasMol92} derived 
the same stochastic \sch equations,  driven
 by the need for efficient 
methods for numerically simulating moderately large quantum systems.
This technique, called Monte-Carlo wavefunction simulations, has been 
applied to great advantage in describing the optical-cooling of a
fluorescent atom
\cite{MolCasDal93,DumZolRit92,MarDumTaiZol93,Mar93,HolMarMarZol96}.
Regardless of the motivation for their use, 
the evolution of systems undergoing 
quantum jumps (and other stochastic quantum 
processes), are known widely as quantum trajectories \cite{Car93b}.

Remarkably, also around the same time as quantum trajectory theory was 
being developed, an entirely different theory of quantum jumps in atomic 
systems was proposed by Teich and Mahler (TM) \cite{TeiMah92}. 
These authors claimed 
that the quantum jumps in their theory also corresponded to photon 
detections. However, it is plain that, except in trivial cases, the 
TM trajectories are different from the quantum 
trajectories from direct photodetection. The internal consistency of 
the TM trajectories has also been criticised by one of us 
\cite{WisMil93c}, but it turns out that there are some subtle issues 
involved here (as will be discussed later). This, 
combined with the simplicity of the TM theory, suggests that it may be 
worth a closer look.

In this paper we use the TM trajectories as a base from which to 
explore quantum jumps  
for the simplest non-trivial atomic system, a  
two-level atom in a strong resonant driving field. 
First (Sec.~II) we summarize  
the TM theory and, following Teich and Mahler, 
derive the TM trajectories for this example. We then identify various 
features of these trajectories, and the interpretation given them by 
Teich and Mahler, and ask the question: can any of these features be 
reproduced by other jump models? Specifically, we examine four other 
jump models: the dressed atom model (Sec.~III), spectral detection 
(Sec.~IV), two-state 
jumps (Sec.~V), and orthogonal jumps (Sec.~VI). The last three models 
derive from quantum trajectories but use detection schemes which are 
progressively harder to implement. The results of these comparisons 
are discussed in Sec.~VII.

\section{The Theory of Teich and Mahler}

\subsection{General Theory}
\label{TMgentheo}

In this section we summarize the theory of Teich and Mahler in 
our own notation. The most general form of Markovian 
master equation for a quantum system is the Lindblad form \cite{Lin76}
\beq \label{me1}
\dot{\rho} = {\cal L}\rho = {\cal L}_{\rm rev}\rho + {\cal L}_{\rm irr}\rho.
\eeq
Here the reversible and irreversible terms are
\bqa
{\cal L}_{\rm rev}\rho = -i[H,\rho] ,\\
{\cal L}_{\rm irr}\rho = \sum_{j}{\cal D}[c_{j}]\rho.
\eqa
Here $H$ is an Hermitian operator, 
the $c_{j}$ are arbitrary operators and ${\cal D}$ is a superoperator 
defined for arbitrary operators $A$ and $B$ as
\beq
{\cal D}[A]B \equiv ABA\dg - \half \{A\dg A,B\}.
\eeq

Say the 
solution of the master equation \erf{me1} is $\rho(t)$. Then 
$\rho(t)$ can be diagonalized at any time as
\beq \label{diag1}
\rho(t) = \sum_{\mu=1}^{D} p_{\mu}(t) 
\ket{\phi_{\mu}(t)}\bra{\phi_{\mu}(t)},
\eeq
where $D$ is the dimension of the Hilbert space of the system.
The time-dependent eigenstates of $\rho(t)$ are always orthogonal to 
one another, and 
so evolve according to a Unitary transformation:
\beq \label{KasH}
\dbd{t}\ket{{\phi}_{\mu}(t)} = -iK(t)\ket{\phi_{\mu}(t)},
\eeq
for some Hermitian operator $K(t)$ which will depend upon $\rho(t)$ 
but not upon $\mu$.
The rate of change of the eigenvalues is thus given by
\bqa
\dbd{t}p_{\mu}(t) %&=& 
%\dbd{t}\bra{\phi_{\mu}(t)}\rho(t)\ket{\phi_{\mu}(t)} \\
&=& \dbd{t}{\rm Tr}[\rho(t) \ket{\phi_{\mu}(t)}\bra{\phi_{\mu}(t)}] \\
%&=& {\rm Tr}[\rho(t) \dbd{t}\ket{\phi_{\mu}(t)}\bra{\phi_{\mu}(t)}] + 
%{\rm Tr}[ \ket{\phi_{\mu}(t)}\bra{\phi_{\mu}(t)}\dbd{t}\rho(t)] \\
&=& -i {\rm Tr}\{ \rho(t) 
[K(t),\ket{\phi_{\mu}(t)}\bra{\phi_{\mu}(t)}]\} \nn \\
&&-i {\rm Tr}\{ \ket{\phi_{\mu}(t)}\bra{\phi_{\mu}(t)}[H,\rho(t) ]
\} \nn \\
&&+ {\rm Tr}[\ket{\phi_{\mu}(t)}\bra{\phi_{\mu}(t)} {\cal L}_{\rm irr}\rho].
\eqa
Using the cyclic properties of the trace operation and the fact that 
$[\ket{\phi_{\mu}(t)}\bra{\phi_{\mu}(t)},\rho(t)]=0$, 
the first two terms vanish. Thus one is left with
\bqa
\dbd{t}p_{\mu}(t) &=&
\bra{\phi_{\mu}(t)}\left[c_{j}\rho(t)c_{j}\dg \right. \nn\\
&& \left. \phantom{\bra{\phi_{\mu}(t)}}-\, \half c_{j}\dg c_{j}\rho(t) - \half \rho(t) c_{j}\dg 
c_{j}\right]\ket{\phi_{\mu}(t)} ,
\eqa
where the Einstein summation convention for 
latin indices is being used. Using \erf{diag1} and 
the completeness relation $1 = \sum_{\nu} \ket{\phi_{\nu}(t)}
\bra{\phi_{\nu}(t)}$,
one obtains finally
\bqa
\dbd{t}p_{\mu}(t) &=& 
\sum_{\nu} \left[ \bra{\phi_{\nu}(t)}c\dg_{j}\ket{\phi_{\mu}(t)}
\bra{\phi_{\mu}(t)}c_{j}\ket{\phi_{\nu}(t)}p_{\nu}(t) \right.\nn\\
&& - \,\left.\bra{\phi_{\mu}(t)}c\dg_{j}\ket{\phi_{\nu}(t)}
\bra{\phi_{\nu}(t)}c_{j}\ket{\phi_{\mu}(t)}p_{\mu}(t)\right]. 
\label{dbdtpmu1}
\eqa

Teich and Mahler interpret the state matrix $\rho(t)$ as pertaining 
to an {\em ensemble} of individual systems. The individual systems, 
they say, are always in one of the states $\ket{\phi_{\mu}(t)}$ 
but jump stochastically between these states.
To derive the rates of these jumps, they rewrite \erf{dbdtpmu1} as
\beq \label{TMj1}
\dot{p}_{\mu}(t) = \sum_{\nu} R_{\mu\nu}(t)p_{\nu}(t) - R_{\nu\mu}(t)p_{\mu}(t),
\eeq
where
\beq \label{Rmunu1}
R_{\mu\nu}(t) = 
\sum_{j}\left|\bra{\phi_{\mu}(t)}c_{j}\ket{\phi_{\nu}(t)}\right|^{2}.
\eeq
In this form it is clear that $R_{\mu\nu}$ can be interpreted as the 
probability per unit time for the system in state 
$\ket{\phi_{\nu}(t)}$ to jump to state $\ket{\phi_{\mu}(t)}$.  
Note that jumps which leave the system in the 
same state also occur, since $R_{\mu\mu}$ is in general non zero.
Between jumps, the system remains in one of the states 
$\ket{\phi_{\mu}(t)}$, and hence changes smoothly in time according 
to \erf{KasH}. It is clear that the TM trajectories do reproduce the master 
equation \erf{me1}, as \erf{TMj1} is equivalent to \erf{me1}.

From the form of \erf{Rmunu1} it appears that the rates
may depend on the way that ${\cal L}_{\rm irr}$ is written. 
The choice of operators $c_{j}$ in the definition of ${\cal L}_{\rm 
irr}$ is not unique; a unitary rearrangement leaves ${\cal L}_{\rm 
irr}$ invariant. That is, 
if we define 
\beq
c_{k}' = U_{kj}c_{j},
\eeq
where $U_{kj}U_{kl}^{*}$ then 
\beq
\sum_{j}{\cal D}[c_{j}] = \sum_{k}{\cal D}[c_{k}'].
\eeq
It turns out that this rearrangement leaves $R_{\mu\nu}$ unchanged 
also:
\beq
\sum_{j}\left|\bra{\phi_{\mu}}c_{j}\ket{\phi_{\nu}}\right|^{2}
= \sum_{k}\left|\bra{\phi_{\mu}}c'_{k}\ket{\phi_{\nu}}\right|^{2},
\eeq
so that the TM trajectories do not depend on the way that 
${\cal L}_{\rm irr}$ is written.

However, it turns out that 
the rates $R_{\mu\nu}$ {\em do} depend on the way that 
${\cal L}$ is split into ${\cal L}_{\rm rev}$ and ${\cal L}_{\rm 
irr}$. It is possible to change ${\cal L}_{\rm rev}$ 
and ${\cal L}_{\rm irr}$ while leaving ${\cal L}$ unchanged by the 
following transformation:
\bqa
c_{j} &\to& c_{j}' = c_{j} + \lambda_{j}, \label{tf1}\\
H &\to& H' = H - \frac{i}{2}[\lambda_{j}^{*}c_{j} - \lambda_{j}c\dg_{j}],
\label{tf2}
\eqa
where the $\lambda_{j}$ are arbitrary c-numbers.
It is simple to see $R_{\mu\nu}$ is {\em not} invariant under this 
transformation. Teich and Mahler do not state what  
$\lambda_{j}$ should be chosen before applying their method; their 
Eq.~(1) already assumes the separation ${\cal L}$ into ${\cal L}_{\rm 
rev}$ and ${\cal L}_{\rm irr}$. One general choice which one might 
make (and which coincides with the choices made by Teich and Mahler in 
the simple systems they consider) is that the $\lambda_{j}$ be such 
that $c_{j}'$ be {\em traceless operators}.

In steady state, $\rho(t)$ is time-independent and we will 
denote it simply
\beq \label{ssdiag}
\rho = \sum_{\mu} p_{\mu} \ket{\phi_{\mu}}\bra{\phi_{\mu}}.
\eeq
In this case the rates are also time-independent and are given by
\beq
R_{\mu\nu} = \sum_{j}
|\bra{\phi_{\mu}}c_{j}\ket{\phi_{\nu}}|^{2}.
\eeq
In what follows we will only consider this stationary stochastic 
evolution.

\subsection{The Two-Level Atom}

Consider an atom with two relevant levels $\{\ket{g},\ket{e}\}$. 
Let  there be a dipole moment between these levels so that the coupling to 
the continuum of electromagnetic field modes in the vacuum state will 
cause the atom to decay  at 
rate $\gamma$. So that the atom does not simply decay to the state 
$\ket{g}$, add driving by a classical field (such as that produced by 
a laser) of Rabi frequency $\Omega$. 
Then the evolution of the atom's state matrix can then be 
described by the master equation (in the interaction picture)
\beq \label{me2}
\dot{\rho} = -i\frac{\Omega}{2}[\sigma_{x},\rho] + \gamma {\cal D}[\sigma]\rho.
\eeq
Here $\sigma = \ket{g}\bra{e}$ and $\sigma_{x}=\sigma+\sigma\dg$.

The stationary stationary solution of \erf{me2} is
\beq
\rho = \frac{\Omega^{2}+\Omega\gamma\sigma_{y} +\gamma^{2}(1-\sigma_{z})/2
}{2\Omega^{2}+\gamma^{2}},
\eeq
where the Pauli matrices have their usual meaning. To simplify 
matters, consider the strong driving limit $\Omega \gg \gamma$. 
Then to first order in $\gamma/\Omega$ we have
\beq
\rho = \frac{\Omega + \gamma\sigma_{y}}{2\Omega}.
\eeq
Diagonalizing this $\rho$ as in \erf{ssdiag} yields the following 
\bqa
\rt{2}\ket{\phi_{1}} &=& \ket{g}-i\ket{e} \label{phi1},\\
\rt{2}\ket{\phi_{2}} &=& \ket{g}+i\ket{e} \label{phi2},\\
p_{1} &=& \frac12\left(1+\frac{\gamma}{\Omega}\right) ,\\
p_{2} &=& \frac12\left(1-\frac{\gamma}{\Omega}\right).
\eqa

Following the choice in \erf{me2} of 
${\cal L}_{\rm irr}=\gamma{\cal D}[\sigma]$, the jump rates can be 
written 
\beq \label{Rmunu2}
R_{\mu\nu} = \gamma 
|\bra{\phi_{\mu}}\sigma\ket{\phi_{\nu}}|^{2}.
\eeq
Substituting the expressions for $\ket{\phi_{1,2}}$ into \erf{Rmunu2} 
gives
\beq
R_{11}=R_{12}=R_{21}=R_{22}=\gamma/4.
\eeq
The fact that $R_{21}=R_{12}$ means that the TM trajectories would 
actually predict $p_{1}=p_{2}=1/2$. To obtain rates which would give 
the correct probabilities it would be necessary to 
determine $\ket{\phi_{1,2}}$ to higher order in $\gamma/\Omega$.

Teich and Mahler associate each jump, including those which 
leave the system state unchanged, with the emission of a photon.
In this case the total rate of photoemission is therefore $\gamma/2$. 
To first order in $\gamma/\Omega$ this agrees with the expression 
from the stationary density operator:
\beq
\gamma\bra{e}\rho\ket{e} = \frac{\gamma}{2}+ O\left( 
\frac{\gamma^{3}}{\Omega^{2}}\right).
\eeq
Photons associated with state-preserving jumps Teich and Mahler 
assign to the central peak of the Mollow triplet, while the 
state-changing emissions they assign to the sidebands. The ratio 
of intensities in the central peak and the sidebands thus agree with those in 
the Mollow triplet \cite{Mol69}.

The  
TM trajectories have a number of interesting characteristics. 
For the purposes of the rest of the paper, there are three in 
particular to which we wish to draw attention: (i) In steady state, the jumps supposedly correspond to photoemissions into the 
	three peaks of the Mollow triplet; (ii) In steady state, the atomic state is always one of 
	two fixed pure states; (iii) At all times, the state after a jump is orthogonal to the 
	one before.   
\label{secfea}

\section{The Dressed Atom Model}

The dressed atom model \cite{CohRey79} can be applied to an arbitrary atomic 
system 
but for simplicity we will consider only the case at hand, the 
resonant two-level atom. The model is based on replacing the 
Hamiltonian in \erf{me2}, in which the driving laser is treated as a 
classical field, with one in which the driving laser 
is treated as a single-mode quantum field. 
Putting in the self-Hamiltonians of the atom and field gives 
(in the rotating-wave approximation)
\begin{equation}
	H=\hbar\omega_0\left(a\dg a + \sigma\dg\sigma \right)
	+ \hbar\frac{g}{2}(a\dg \sigma + a \sigma\dg),
\end{equation}
where $a$ is the annihilation operator for the driving field and
$g$ is the dipole coupling constant also known as the one-photon 
Rabi frequency. This Hamiltonian has eigenstates
\beq  \label{defdrest}
\rt{2}|n,\pm\rangle = |n\rangle |g\rangle \pm |n-1\rangle 
|e\rangle  ,
\eeq
where $|n\rangle$ are number states of the driving field.
These eigenstates are known as dressed 
states of the atom, as opposed to $\{\ket{e},\ket{g}\}$, the bare 
atomic energy states. They have energies
\beq E_{n,\pm} = \hbar ( n\omega_0 \pm \sqrt{n}\, g/2). \label{ladder}
\eeq

 For large coherent driving field and small
 coupling constant $g$ the classical approximation is valid and one can 
 replace $\sqrt{n} g$ by $\sqrt{\bar{n}} g = \Omega$.
 Then, for $n\sim\bar{n}$, the ladder of energy eigenstates 
 (\ref{ladder}) will consist of 
 pairs of closely-spaced rungs, with an inter-pair separation of 
 $\hbar\omega_0$, and an intra-pair separation of $\hbar\Omega$.  
 
 Now one can interpret the Mollow triplet in terms of 
spontaneous-emission-induced transitions between these 
stationary states. If the dressed atom 
is in one of the states $|n,\pm\rangle$, it can spontaneously 
emit a photon and drop down a rung on 
the ladder. If it drops to $|n-1,\pm\rangle$ (that is,
 the atom effectively remains in the 
same state), then the change in energy 
of the dressed atom is $\hbar\omega_0$ and 
so the frequency of the emitted photon must be $\omega_0$
 --- in the central peak of the 
triplet. If the atom changes state via a transition to 
$|n-1,\mp \rangle$, then the 
frequency of the emitted photon must be $\omega_0 \pm 
\Omega$ --- in the sideband peaks. 

The rates of these transitions are calculated according to the same 
formula as in Teich and Mahler's scheme. For example, if the system 
were in the dressed state $\ket{n,+}$ then the probability of it to 
spontaneously emit and make a transition to the dressed state 
$\ket{n-1,-}$ is 
\beq
R_{-+} = \gamma \left| \bra{n-1,-}\sigma\ket{n,+} \right|^{2}.
\eeq
Substituting this expression into \erf{defdrest}, and doing similarly for 
the other transitions, gives
\beq
R_{-+}=R_{++}=R_{+-}=R_{++}=\gamma/4.
\eeq
That is, all of the transition rates are equal. A strict ordering of 
transitions will also occur, ensuring that a high frequency photon 
must be emitted between emissions of low frequency photons, and {\em 
vice versa}. 

All of the features of the dressed-state model discussed above agree 
with those of the TM theory. Teich and Mahler even go so far as to 
call their theory a ``generalization of the dressed state picture.'' 
\cite{TeiMah92} 
However, in one crucial respect the two theories are completely at 
odds: the state of the atom. In TM theory it jumps between the states 
$\ket{\phi_{1}}$ and $\ket{\phi_{2}}$ of Eqs.~(\ref{phi1}) and 
(\ref{phi2}). 
These are eigenstates of $\sigma_{y}$. By contrast, for a coherent 
driving field the dressed states of the atoms are
\beq
\sum_{n=0}^{\infty} e^{-|\alpha|^{2}/2}\frac{\alpha^{n}}{\rt{2(n!)}} 
\left(|n\rangle |g\rangle \pm |n-1\rangle 
|e\rangle \right).
\eeq
For $|\alpha|$ large the field and atom states are very nearly 
not entangled, and the state of the atom is very close to a pure 
state $\ket{\pm}$. For $\alpha$ real, as required to make the 
replacement
\beq
\frac{g}{2}(a\dg \sigma + a \sigma\dg) \to 
\frac{g\alpha}{2}(\sigma + \sigma\dg),
\eeq
so as to yield the master equation (\ref{me2}) with $\Omega=g\alpha$, 
these pure states are defined by
\beq \label{ds}
\rt{2}\ket{\pm} = \ket{g}\pm \ket{e}.
\eeq
These are eigenstates of $\sigma_{x}$. Thus the atomic states in the 
dressed atom model are as different as it possible to be (on the Bloch 
sphere) from the atomic states in the TM model.

\section{Spectral Detection}

The disagreement between the dressed state model and the TM model, 
both of which claim to describe emission into the three peaks of the 
Mollow triplet, suggest that it would be worthwhile investigating 
such emissions by a third, independent theory. The theory we use in 
this section is that of quantum trajectories, developed initially by 
Carmichael \cite{Car93b} and subsequently by many other authors.
This theory is essentially an application of quantum measurement
theory of continuously monitored systems, so we will briefly review 
this theory. We will consider only {\em efficient} measurements
in which no information is lost. In the optical context, this 
requires complete collection of the emitted light and unit efficiency 
photodetectors.

\subsection{Quantum Trajectories}

The aim of quantum measurement theory is, given the initial state of the system,
to be able to specify the
probability of a particular measurement result, and the state of
the system immediately following this result. Say the
measurement result is $\alpha$, a random variable which will
be assumed discrete for convenience. Then both the probability and
the conditioned state can be found from the set of operators
$\{\Omega_\alpha(T)\}$. Here $T$ is the 
duration of the measurement and the operators $\Omega_{\alpha}(T)$ 
are arbitrary, with one condition, 
\begin{equation}  \label{complet}
\sum_\alpha \Omega_\alpha ^\dagger(T) \Omega_\alpha(T) =1,
 \end{equation} 
where the sum is over all possible results. This is known as the
completeness condition \cite{Gar91}.

The probability for obtaining a particular result $\alpha$ is found
from the measurement operator by
\begin{equation} \label{prob}
P_\alpha = {\rm Tr}[\tilde{\rho}_\alpha(t+T)],
\end{equation}
where 
\begin{equation} \label{unnorm}
\tilde{\rho}_\alpha(t+T) = \Omega_\alpha(T) \rho(t) \, \Omega_\alpha 
^\dagger(T)
\end{equation}
is an unnormalized density operator, where $\rho(t)$ is the density
operator at the beginning of the measurement. The state of the system 
conditioned on the result $\alpha$ is simply given by 
 \begin{equation} \label{condit} 
\rho_\alpha(t+T) = \tilde{\rho}_\alpha(t+T) /P_\alpha.
 \end{equation}
This economy of theory is a consequence of a more fundamental
notion of probability relating to projectors in Hilbert space
\cite{Hol82}. If the initial
state of the system is pure ($\rho =
|\psi\rangle\langle\psi|$), then the unnormalized
conditioned state is obviously
\begin{equation}
|\tilde{\psi}_\alpha(t+T) \rangle  = \Omega_\alpha(T) |\psi(t)\rangle .
\end{equation}
If the measurement is performed, but the
result ignored, then the new state of the system will be mixed in
general and cannot be represented by a state vector. It is equal to the
sum of the conditioned density operators (\ref{condit}), weighted by
the probabilities (\ref{prob})
 \begin{eqnarray} 
\rho(t+T) &=&\sum_\alpha P_\alpha \rho_\alpha(t+T) \nonumber \\ 
&=& \sum_\alpha \Omega_\alpha(T) \rho(t)\, \Omega_\alpha ^\dagger(T). 
\end{eqnarray} 
It is easy to verify from the completeness condition (\ref{complet}) 
that ${\rm Tr}[\rho(t+T)]=1$, as required by conservation of probability.
 
Continuous measurement theory can now be cast as 
a special case of quantum measurement theory where a 
constant measurement interaction allows successive measurements, 
the duration of each being infinitesimal. If the 
state matrix at time $t$ is $\rho(t)$, then the unnormalized 
conditioned density operator after the measurement in the interval 
$[t,t+dt)$ is denoted
\begin{equation}
\tilde{\rho}_\alpha (t+dt) = \Omega_\alpha(dt) \rho(t) \, 
\Omega_\alpha ^\dagger (dt),
\end{equation}
where the operators $\Omega_\alpha(dt)$ are arbitrary as yet. The 
unconditioned infinitesimally evolved state matrix is then
\begin{equation}
\rho(t+dt) = \sum_\alpha  \Omega_\alpha(dt) \rho(t) \, \Omega_\alpha 
^\dagger (dt).
\end{equation}
This represents the evolution of the system, ignoring the measurement 
results. If the $\Omega_\alpha(dt)$ are time-independent, this 
nonselective evolution is obviously Markovian (depending only on the 
state of the system at the start of the interval).

As noted in Sec.~\ref{TMgentheo}, the most general form of
Markovian equation of motion for the density operator of a system 
is a master equation of the Lindblad form \cite{Lin76}. Consider a 
special case of \erf{me1} where there is a single Lindblad operator 
$c$. Then, by inspection, there 
are only two necessary measurement results (say 0 and 1), with 
corresponding operators
\begin{eqnarray}
\Omega_1(dt) &=& \sqrt{dt}\, c , \label{O1}\\
\Omega_0(dt) &=& 1 - \left[iH + \case{1}/{2}c^\dagger c \right] dt. 
\label{O0}
\end{eqnarray}
The two unnormalized conditioned density operators are
\begin{eqnarray}
\tilde{\rho}_1(t+dt) &=& dt\, c\rho c^\dagger ,\\
\tilde{\rho}_0(t+dt) &=& \rho + dt \left( -i[H,\rho] - \case{1}/{2} 
\{ c^\dagger c ,\rho \} \right).
\end{eqnarray}
It is easy to see that
\begin{equation}
\rho(t+dt) = \tilde{\rho}_1(t+dt) + \tilde{\rho}_0(t+dt) = \rho + dt 
\dot{\rho},
\end{equation}
where $\dot{\rho}$ is given by the master equation (\ref{me1}). 

Evidently, almost all infinitesimal intervals yield the measurement 
result 0. Upon such result, the system state evolves infinitesimally 
(but not unitarily). Whenever the result 1 is obtained, 
however, the system state changes by a finite operation. This 
discontinuous change can be justifiably called a {\em quantum jump}, 
and the measurement event a {\em detection}. 
As with all efficient measurements, if the initial state of the system is 
pure, then the 
conditioned state of the system will remain pure. The stochastic 
evolution of such a conditioned state has been called a {\em quantum 
trajectory} by Carmichael \cite{Car93b}, as discussed in the 
introduction. 

Now consider the application of continuous quantum measurement theory 
to the damped, driven 
two-level atom which is the subject of this paper. The master equation for 
this system is given by \erf{me2}, and this can be most simply unraveled 
using the two measurement operators 
\bqa
\Omega_0(dt) &=& 1 - i\frac{\Omega}{2}\sigma_{x}dt - 
\frac{\gamma}{2}\sigma\dg \sigma dt, \\
\Omega_1(dt) &=& \rt{\gamma dt}\sigma.
\eqa
In this unraveling, the rate of jumps is $\gamma \ip{\sigma\dg \sigma}$, 
which is identical to the rate of spontaneous emissions. Thus 
this unraveling corresponds to direct detection (by a photodetector) of 
all of the light emitted by the atom. Indeed, the detection 
operator $\Omega_{1}(dt)$ is proportional to the part of the field 
radiated by the atom.

It is evident that under this detection scheme, the atomic state 
immediately following a jump is always the ground state $\ket{g}$. Thus 
these quantum jumps are quite different from those in either the dressed 
state or the TM model. This is not surprising, as the photodetector does 
not distinguish photons of different frequencies, as required in these 
other models. 

In order to distinguish photons of different frequency it is necessary to 
use optical filters in the detection apparatus prior to photodetection. 
There are a number of ways to describe such a detection scheme. One is 
to use a non-Markovian evolution equation for the system, 
as done by Jack, Collett and Walls \cite{JacColWal98}. However, this 
has the added complication that the system is represented not by a 
single state vector but by a 
sequence of state vectors (which in principle is infinite). 
Also, this formalism does not 
directly give the state of the system at the time $t$ of the 
measurement, but rather the {\em retrodicted} state \cite{PegBar99} 
at time $t-T_{m}$. Here $T_{m}$ is the time required by the filtering 
process. In this work we adopt a different procedure in which the 
filters are described as quantum optical systems (cavities). In our 
formalism only  a single state vector is required, but it exists in 
the enlarged Hilbert space of the system (the two-level atom) plus 
filters. This state vector applies to the system plus filters at the 
moment of measurement (ignoring the propagation time for light between 
the various elements). Finally, our method is amenable to an 
approximate analytic solution in a suitable limit.

\subsection{Cascaded quantum systems}

In order to describe auxiliary quantum systems (filters) as part of 
the detection scheme it is necessary to use 
``cascaded systems theory'' as it has been called by Carmichael \cite{Car93a}. 
 Cascaded 
systems are different from coupled systems, because the interaction only 
goes one way. That is to say, the first system influences the second, but 
not {\em vice versa}. One mechanism for achieving the required 
unidirectionality is the Faraday isolator which utilizes Faraday 
rotation and polarization-sensitive beam splitters. This is most 
practical for the case of cavities, where the output is a beam of light. 
In principle this technique could be applied to the radiation of an 
atom as well, if the atom were placed at the focus of a parabolic 
mirror so as to produce an output beam, as shown in Fig.~1.

A quantum 
theoretical treatment which incorporates this spatial symmetry 
breaking at the level of the Hamiltonian has been given by 
Gardiner \cite{Gar93}. If the propagation time between the source 
system the and driven system is negligible then a master equation for 
both systems may be derived. This result was obtained simultaneously 
by Carmichael \cite{Car93a}, who used quantum trajectories to illustrate 
the nature of the process. Since we wish to describe the monitoring 
of the outputs of the filter, we will follow Carmichael's approach.

Let the first system be a two level atom obeying the usual master equation
\beq
\dot{\rho} = -i\left[ ({\Omega}/{2})\sigma_{x} ,\rho\right] + 
{\cal D}[\rt{\gamma}\sigma]\rho. \label{me5}
\eeq
As noted above, the field radiated by this atom is represented by the 
operator $\rt\gamma \sigma$.  
Now say this field (plus the accompanying electromagnetic 
vacuum fluctuations) impinges upon the front mirror of 
a Fabry-Perot etalon of linewidth 
$2\Gamma$. That is to say, the intensity decay rates through the front and 
rear mirror are both $\Gamma$. Then the master equation for the 
state matrix $W$ for the combined system is
\bqa
\dot{W} &=& -i\left[ ({\Omega}/{2})\sigma_{x} +
\omega_{a}a\dg a + i\rt{\gamma\Gamma}(\sigma\dg a - \sigma 
a\dg),W\right] 
\nn \\ 
&& + \, {\cal D}[\rt{\gamma}\sigma+\rt{\Gamma}a]W + 
{\cal D}[\rt{\Gamma}a]W ,\label{casc1}
\eqa
where $\omega_{a}$ is the {\em detuning} of the relevant mode of the etalon 
(relative to the atom and its resonant driving field) and 
$a$ is its annihilation operator.

It can be verified from \erf{casc1} that tracing over 
the cavity mode $a$ yields \erf{me5} for the atom alone. That is to 
say, the filter does not directly affect the atom, which is as desired. The 
apparent coupling  term in the Hamiltonian in 
\erf{casc1} is canceled by the interference in the irreversible term 
with Lindblad operator $\rt{\gamma}\sigma+\rt{\Gamma}a$. This operator 
represents the radiated field from the front of the resonator. It is 
the Faraday isolator or equivalent mechanism which prevents the 
interaction of this field with the atom. The 
second Lindblad operator $\rt{\Gamma}a$ 
represents the field radiated from the rear 
of the resonator.

Since the filter produces two output fields (that passed and that 
rejected), monitoring the system requires two photodetectors. Hence 
the system will now be described by three measurement operators, 
$\Omega_{1}(dt)$ corresponding to the detection of a ``rejected'' photon 
(off the front of the etalon), $\Omega_{a}(dt)$ corresponding to a 
``passed'' photon (from the rear), and $\Omega_{0}(dt)$ corresponding to 
no detection of a photon in the interval $[t,t+dt)$. These operators 
are given by
\bqa
\Omega_{1}(dt) &=& 
\rt{\gamma dt}\sigma+\rt{\Gamma dt}a,\\
\Omega_{a}(dt) &=& \rt{\Gamma dt}a ,\\
\Omega_{0}(dt) &=& 1 - idt\left(
\omega_{a}a\dg a + \frac{\Omega}{2}\sigma_{x}\right) \nn \\
&&\phantom{1} - dt\left( \frac{\gamma}{2}\sigma\dg \sigma + \Gamma a\dg a + 
\rt{\gamma\Gamma}a\dg \sigma\right).
\eqa
It is easy to verify that 
\beq
\sum_{\alpha = 0,1,a} \Omega_{\alpha}(dt)\dg W \Omega_{\alpha}(dt)
 = \dot{W}dt,
 \eeq
where $\dot{W}$ is given by \erf{casc1}.

\subsection{Two-Level Atom with One Filter}

We wish to consider now the specific case where the filter cavity 
is designed so as to pass photons from the high frequency peak of the 
Mollow triplet, while rejecting photons from the middle and low 
frequency peaks. In the limit $\Omega \gg \gamma$ (which is required 
for the three peaks to be well separated), the upper peak is centered 
at frequency $\omega_{0} + \Omega$. Thus we choose the detuning of 
the etalon to be $\omega_{a} = \Omega$. 
The two measurement operators $\Omega_{1}(dt)$ and 
$\Omega_{a}(dt)$ are unchanged, while $\Omega_{0}(dt)$ then becomes
\bqa
\Omega_{0}(dt) &=& 1 - idt \frac{\Omega}{2} \left(\sigma_{x}+2 
a\dg a\right) \nn \\
&&\phantom{1} - dt\left( \frac{\gamma}{2}\sigma\dg \sigma + \Gamma a\dg a + 
\rt{\gamma\Gamma}a\dg \sigma\right).
\eqa

Now in order to pass almost all of the high frequency photons but 
reject almost all of the middle and low frequency photons we require 
a filter bandwidth $2\Gamma$ satisfying $\Omega \gg \Gamma \gg \gamma$. 
In this limit the cavity relaxes much faster than the rate of 
spontaneous emission by the atom. Since on the cavity time scale, 
photodetections are infrequent events, it makes sense to consider the 
basis which diagonalizes the no-jump operator $\Omega_{0}(dt)$. 
Because this operator is not normal (that is, it does not satisfy 
$[\Omega_{0}(dt)\dg,\Omega_{0}(dt)]=0$), its eigenstates are not 
orthogonal. Nevertheless they form a complete basis and to zeroth order 
in $(\gamma/\Omega)$ and $(\gamma/\Gamma)$ they are 
orthonormal. The exact expressions for the eigenstates are
\beq \label{defS1}
\ket{S_{n}^{\pm}} = \sum_{j=n}^{\infty}\left( h_{nj}^{\pm}\ket{h}\ket{j}
+ l_{nj}^{\pm}\ket{l}\ket{j} \right).
\eeq
Here $\ket{h}$ and $\ket{l}$ are atomic states defined by
\bqa
\ket{h} &=& \mu\ket{g}+\nu\ket{e}, \label{keth}\\
\ket{l} &=& \nu\ket{g} - \mu\ket{e} \label{ketl},
\eqa
where 
\beq
\frac{\nu}{\mu} = {\rt{1-\frac{\gamma^{2}}{4\Omega^{2}}} - 
\frac{i\gamma}{2\Omega}}.
\eeq
Note that these atomic states are not exactly orthogonal. However, with 
an error of order $(\gamma/\Omega)^{2}$ they are equal to the dressed 
states of \erf{ds}:
\bqa
\ket{h} = \ket{+} + O(\gamma^{2}/\Omega^{2}), \\
\ket{l} = \ket{-} + O(\gamma^{2}/\Omega^{2}),
\eqa
and hence are very nearly orthogonal. In \erf{defS1} the states 
$\ket{j}$ are eigenstates states of $a\dg a$. The coefficients 
$h^{\pm}_{nj}, l^{\pm}_{nj}$ are defined by the recurrence relations
\begin{eqnarray}
h_{nj}^\pm &=& \frac{\sqrt{\gamma\Gamma j}}{\lambda_h + \lambda_a j - 
\sigma_n^\pm} \left[ q h_{n,j-1}^\pm + (p-1)l_{n,j-1}^\pm \right] ,
\label{4.29} \\
l_{nj}^\pm &=& \frac{\sqrt{\gamma\Gamma j}}{\lambda_l + \lambda_a j - 
\sigma_n^\pm} \left( h_{n,j-1}^\pm p - q l_{n,j-1}^\pm \right)  ,
\label{4.30}
\end{eqnarray}
for $j>n$ where 
\beq
p = \frac{\nu^{2}}{\mu^{2}+\nu^{2}} \;,\;\;
q = \frac{\mu\nu}{\mu^{2}+\nu^{2}},
\eeq
and the initial conditions for each recurrence chain are
\begin{eqnarray}
 h_{nn}^+ = 1 \;&,&\;\;  l_{nn}^+=0 ,\label{4.31} \\
 h_{nn}^- = 0 \;&,&\;\;  l_{nn}^-=1 .\label{4.32}
\end{eqnarray}
In Eqs.~(\ref{4.29}) and (\ref{4.30}),
\begin{eqnarray}
\sigma_n^{+} &=& \lambda_h + n \lambda_a ,\\
\sigma_n^{-} &=& \lambda_l + n \lambda_a ,
\end{eqnarray}
and
\bqa
\lambda_h &=& -\frac{\gamma}{4} - \frac{i}{2}\sqrt{ 
\Omega^2-\frac{\gamma^{2}}{4}}, \label{lamh} \\
\lambda_{a} &=& -\Gamma -i\omega_{a} = -\Gamma -i\Omega ,
\label{lama}\\
\lambda_{l} &=& -\frac{\gamma}{4} + \frac{i}{2}\sqrt{ 
\Omega^2-\frac{\gamma^{2}}{4}} .\label{laml}
\eqa
As well as appearing in the recurrence relations, the 
$\sigma_{n}^{\pm}$ define the eigenvalues:
\beq
\Omega_0(dt)\ket{S_n^\pm} = \left( 1 + \sigma_n^\pm  dt \right) 
\ket{S_n^\pm}. \label{eigval1}
\eeq

Before proceeding further we make the assumption that terms of 
order $(\gamma/\Omega)^{2}$ may be ignored. This will be justified 
later. Under this assumption $\mu=\nu$ and $p=q=\frac12$ so that
\beq \label{defS2}
\ket{S_{n}^{\pm}} = \sum_{j=n}^{\infty}\left( h_{nj}^{\pm}\ket{+}\ket{j}
+ l_{nj}^{\pm}\ket{-}\ket{j} \right),
\eeq
where
\begin{eqnarray}
h_{nj}^\pm &=& \frac{\sqrt{\gamma\Gamma j}}{\lambda_h + \lambda_a j - 
\sigma_n^\pm} \left( h_{n,j-1}^\pm - l_{n,j-1}^\pm \right)/2 ,
\label{4.29a} \\
l_{nj}^\pm &=& \frac{\lambda_h + \lambda_a j - 
\sigma_n^\pm}{\lambda_l + \lambda_a j - 
\sigma_n^\pm} h_{nj}  ,
\label{4.30a}
\end{eqnarray}
where
\bqa
\lambda_h &=& -\frac{\gamma}{4} - \frac{i\Omega}{2}, \label{lamh2} \\
\lambda_{a} &=&  -\Gamma -i\Omega ,
\label{lama2}\\
\lambda_{l} &=& -\frac{\gamma}{4} + \frac{i\Omega}{2} .\label{laml2}
\eqa

The next assumption we make is that we need consider only the two 
states $\ket{S_{0}^{\pm}}$. This is based on the observation that the 
real part of $\sigma_{n}^{\pm}$ is $\gamma/4+n\Gamma$. This means that 
if the state is prepared in a superposition of states $\ket{S_{n}^{\pm}}$,
the decay rate for the component with $n > 0$ is much faster than for 
that with $n=0$, since $\Gamma \gg \gamma$. Thus, given that no 
detection occurs, the system will soon find itself in the subspace 
spanned by the states $\ket{S_{0}^{\pm}}$. We will return later to the 
question of how much error is introduced by this approximation. 
Meanwhile, solving the recurrence relations (\ref{4.29a}), 
(\ref{4.30a}) yields
\bqa
\ket{S_{0}^{+}} &=& \ket{+}\ket{0} - 
\frac{\sqrt{\gamma}}{2\sqrt{\Gamma}}\ket{-}\ket{1} + 
\frac{i\sqrt{\Gamma\gamma}}{2\Omega}\ket{+}\ket{1} ,\\
%&& + \, \frac{-i\rt{2}\gamma}{4\Omega}\left( \ket{-}\ket{2} + 
%\frac{1}{2}\ket{+}\ket{2}\right), \\
\ket{S_{0}^{-}} &=& \ket{-}\ket{0} - 
\frac{i\sqrt{\Gamma\gamma}}{4\Omega}\ket{+}\ket{1} -
\frac{i\sqrt{\Gamma\gamma}}{2\Omega}\ket{-}\ket{1} ,
\eqa
where the terms ignored with $2$ or more photons in the cavity of order
$\gamma/\Omega$.
Note that when the system is in state $\ket{S_{0}^{\pm}}$ the atom is 
substantially in the dressed state $\ket{\pm}$. 
%This suggests that the 
%TM model will compare unfavourably to the dressed state model in 
%describing spectral detection.

Consider the case when the total system is in state 
$\ket{S_{0}^{+}}$. It will remain in this state until a detection 
occurs. If a photon passed through the cavity is detected, 
the new state will be 
proportional to
\bqa
\Omega_{a}(dt)\ket{S_{0}^{+}} &\propto& \ket{-}\ket{0} - 
\frac{i\Gamma}{\Omega}\ket{+}\ket{0} + 
O\left(\frac{\rt{\gamma\Gamma}}{\Omega}\right)\ket{1} \\
&=& \ket{S_{0}^{-}} - \frac{i\Gamma}{\Omega}\ket{S_{0}^{+}}
+ O\left(\frac{\rt{\gamma\Gamma}}{\Omega}\right)\ket{1} .
\eqa
That is, the detection of a high-frequency photon transfers the atomic 
state from being predominantly in the high energy dressed state 
$\ket{+}$ to being predominantly in the low energy dressed state $\ket{-}$. 
This is in line with the simple dressed state model. Moreover, if 
instead a 
rejected photon (from the middle or lower peak) is detected, the new 
state is proportional to 
\bqa
\Omega_{1}(dt) \ket{S_{0}^{+}} &\propto& \ket{+}\ket{0} 
+ O\left(\frac{\rt{\gamma\Gamma}}{\Omega}\right)\ket{1} \\
&=& \ket{S_{0}^{+}} 
+ O\left(\frac{\sqrt{\gamma}}{\sqrt{\Gamma}}\right)\ket{1} .
\eqa
That is, the atomic state is substantially unchanged. This is again 
as expected from the dressed state theory, as a low frequency photon 
from an atom in the dressed state $\ket{+}$ would be impossible, and a 
resonant frequency photon would leave the atomic state unchanged.

The picture so far from this detection scheme is remarkably close to 
the dressed state model. However, the correspondence breaks down when 
we consider the system initially in the state $\ket{S^{-}_{0}}$. Then 
when a passed photon is detected the new state is 
\beq
\Omega_{a}(dt)\ket{S_{0}^{-}} \propto \frac{1}{2}\ket{+}\ket{0} + 
\ket{-}\ket{0} + 
O\left(\frac{\sqrt{\gamma}}{\sqrt{\Gamma}}\right)\ket{1}, 
\eeq
which is not close to either dressed state. Similarly if a rejected 
photon is detected the new state is
\beq
\Omega_{1}(dt)\ket{S_{0}^{-}} \propto \ket{g}\ket{0} + 
O(\Gamma/\Omega). 
\eeq
which is an equal superposition of the two dressed states. 

The reason 
for this failure of the dressed state model is that the detection 
scheme does not distinguish between photons in the central and lower 
peak of the Mollow triplet. It might be thought that tuning the 
cavity to the central frequency $\omega_{0}$ could solve this problem, 
as then a passed photon would leave the atom in the same dressed 
state, while a rejected photon would swap the atom from one dressed 
state to the other. This does work for short times, as shown in 
Fig.~2, if the atom starts in a dressed state. 
However, because there is no way of distinguishing between 
high and low frequency photons, errors accumulate and soon the 
experimenter could not tell which dressed state the atom is in. 
In theory, the atom is still approximately in a pure state, but to 
know what pure state it is in would require timing of the detections 
to time scales less than $\Omega^{-1}$. This would be difficult in 
practice and is also counter to the 
spirit of the dressed state model. 

To attempt to properly replicate the dynamics of the dressed state 
model (or, perhaps, the TM model) it is necessary to distinguish all 
three peaks of the Mollow triplet. This requires two 
filters and is investigated in the following section.

\subsection{Two-Level Atom with Two Filters}

Consider the set up shown in Fig.~3 with two cascaded filter 
cavities with annihilation operators $a$ and $b$. The master equation 
for this system is
\bqa
\dot{W} &=& -i\left[ ({\Omega}/{2})\sigma_{x} + 
\omega_{a}a\dg a + \omega_{b}b\dg b ,W \right]\nn  \\
&& 
- \, i\rt{\Gamma}[\rt{\gamma}(i\sigma\dg a - i\sigma a\dg),W]\nn \\
&& - \,i\rt{\Gamma}[i(\rt{\gamma}\sigma+\rt{\Gamma}a)\dg b 
- i(\rt{\gamma}\sigma+\rt{\Gamma}a)b\dg,W]
\nn \\ 
&& + \, {\cal D}[\rt{\gamma}\sigma+\rt{\Gamma}a]W + 
{\cal D}[\rt{\Gamma}a]W\nn \\
&& + \, {\cal D}[\rt{\gamma}\sigma+\rt{\Gamma}a+\rt{\Gamma}b]W .\label{casc2}
\eqa
Here we have chosen the bandwidth of the second cavity to equal that 
of the first, $2\Gamma$. 

In this case we have three detection events. If we choose 
$\omega_{a}=+\Omega$ (as before) and 
$\omega_{b}=-\Omega$, then photons passed 
by filters $a$ and $b$ will be from the high and low sidebands 
respectively and 
those rejected by both cavities will fall in the central peak.
The four measurement operators are, in the usual interaction picture,
\begin{eqnarray}
\Omega_{1}(dt) &=& 
\rt{\gamma dt}\sigma+\rt{\Gamma dt}a+\rt{\Gamma dt}b,\\
\Omega_{a}(dt) &=& \rt{\Gamma dt}a ,\\
\Omega_{b}(dt) &=& \rt{\Gamma dt}b ,\\ 
\Omega_0(dt) &=& 1 - idt \frac{\Omega}{2} \left(\sigma_{x}+2 
a\dg a-2 b\dg b\right) \nn \\
&&\phantom{1} - dt\left( \frac{\gamma}{2}\sigma\dg \sigma + \Gamma a\dg 
a + \Gamma b\dg b \right) \nn \\
&&\phantom{1} - dt\left(\rt{\gamma\Gamma}a\dg \sigma + \Gamma b\dg 
a + \rt{\gamma\Gamma}b\dg \sigma \right). 
\end{eqnarray}
Note that detection of a  rejected photon now involves field 
amplitudes emitted from the atom and both cavities.

To attack the problem we again find the eigenstates of 
$\Omega_{0}(dt)$. For $n,m$ natural numbers these are given by  
\begin{equation}
\ket{S_{nm}^\pm} = \sum _ {j=0}^\infty \sum_{k=0}^\infty \left( 
h_{nmjk}^\pm\ket{h}\ket{j}\ket{k} + 
l_{nmjk}^\pm\ket{l}\ket{j}\ket{k} \right).
\label{4.50}
\end{equation}
The recurrence relationship for $h_{nmjk}^\pm$ and 
$l_{nmjk}^\pm$ is slightly more complicated than for the single 
cavity case.  
\begin{eqnarray}
h_{nmjk}^\pm &=& \left( \lambda_h + 
\lambda_a j + \lambda_b k - \sigma_{nm}^\pm \right)^{-1} \nn \\
&& \times\, \left\{ \rt{\gamma\Gamma j}\left[q h_{nm,j-1,k}^\pm 
+ (p-1) l_{nm,j-1,k}^\pm\right] \right. \nn \\
&& \phantom{\times\, } +\, \rt{\gamma\Gamma k} \left[ q 
h_{nmj,k-1}^\pm + (p-1)l_{nmj,k-1}^\pm\right] 
\nn \\
&& \phantom{\times\, }\left. +\,
\Gamma \rt{k(j+1)} h_{nm,j+1,k-1}^\pm \right\}, \label{4.53} \\
l_{nmjk}^\pm &=& \left( \lambda_l + 
\lambda_a j + \lambda_b k - \sigma_{nm}^\pm \right)^{-1} \nn \\
&& \times\, \left\{ \rt{\gamma\Gamma j}\left[p h_{nm,j-1,k}^\pm 
-q l_{nm,j-1,k}^\pm\right] \right. \nn \\
&& \phantom{\times\, } +\, \rt{\gamma\Gamma k} \left[ p 
h_{nmj,k-1}^\pm -q l_{nmj,k-1}^\pm\right] 
\nn \\
&& \phantom{\times\, }\left. +\,
\Gamma \rt{k(j+1)} l_{nm,j+1,k-1}^\pm \right\}.  \label{4.54}
\end{eqnarray}
Here $\lambda_{a}=-\Gamma-i\Omega$ as before while $\lambda_{b} = 
-\Gamma +i\Omega$, and the eigenvalues of $\Omega_{0}(dt)$ are 
$1+\sigma_{nm}^{\pm}(dt)$ where
\bqa
\sigma^{+}_{nm} &=& \lambda_{h} + \lambda_{a} n + \lambda_{b} m, \\
\sigma^{-}_{nm} &=& \lambda_{l} + \lambda_{a} n + \lambda_{b} m .
\eqa
The boundary conditions for the recurrence relations are
\begin{eqnarray}
 h_{nmnm}^+ = 1 \;&,&\;\;  l_{nmnm}^+=0 ,\label{4.51} \\
 h_{nmnm}^- = 0 \;&,&\;\;  l_{nmnm}^-=1 .\label{4.52}
\end{eqnarray}  
Note that there is an asymmetry in $j$ and $k$ in Eqs.(\ref{4.53}) 
and (\ref{4.54}) due to the ordering of the cavities. This is also 
manifest in the the range of $j$ and $k$ for a given $n$ and $m$.  
Starting at $k=m$, $j$ should range from $j=n+1$ to $\infty$, while
 $k$ 
is then continually incremented, and for each $k$ value, $j$ should 
run from max$(n-k+m,0)$ to infinity. 

As in the single cavity case, we are interested in the longest-lived 
states $\ket{S_{00}^{\pm}}$. Also, we make the 
the same approximations stemming 
from the limits $\Omega \gg \Gamma \gg \gamma$. Under these 
approximations we find the following 
expressions (where the $00$ subscript has been omitted)
\bqa
\ket{S^{+}} &=& \ket{+}\ket{00} - 
\frac{\sqrt{\gamma}}{2\sqrt{\Gamma}}\ket{-}\ket{10} \nn \\
&& +\, \frac{i\sqrt{\Gamma\gamma}}{2\Omega}\ket{+}\left(\ket{10} - 
\ket{01}\right) \nn \\
&& -\, \frac{\gamma}{8\Gamma}\ket{+}\ket{11} + 
O\left(\frac{\gamma}{\Omega}\right) + 
O\left(\frac{\gamma^{3/2}}{\Gamma^{3/2}}\right) , \label{466}\\
\ket{S^{-}} &=& \ket{-}\ket{00} + 
\frac{\sqrt{\gamma}}{2\sqrt{\Gamma}}\ket{+}\ket{01} \nn \\
&& +\, \frac{i\sqrt{\Gamma\gamma}}{4\Omega}\left(2\ket{-}\ket{01} - 
2\ket{-}\ket{10} - \ket{+}\ket{10}\right) \nn \\
&& -\, \frac{\gamma}{8\Gamma}\ket{-}\ket{11} + 
O\left(\frac{\gamma}{\Omega}\right) + 
O\left(\frac{\gamma^{3/2}}{\Gamma^{3/2}}\right) . \label{467}
\eqa
Here the omitted states of order $\gamma/\Omega$ have two photons in 
one cavity and none in the other, while those of order 
$(\gamma/\Gamma)^{3/2}$ have two in one and one in the other. 
Note the asymmetry in the terms of order $\rt{\Gamma\gamma}/\Omega$ 
due to the ordering of the cavities.

Now imagine the total system is in state $\ket{S^{+}}$. It will 
remain in that state until a detection occurs. The new state conditioned 
on the detection of a photon passed by cavity $a$ is
\bqa
\Omega_{a}(dt)\ket{S^{+}} &\propto& 
 \ket{-}\ket{00} + 
\frac{-i\Gamma}{\Omega}\ket{+}\ket{00}\nn \\
&&+\, 
\frac{\sqrt{\gamma}}{4\sqrt{\Gamma}}\ket{+}\ket{01} + O\left( 
\frac{\sqrt{\Gamma\gamma}}{\Omega}\right). \label{morcom}
\eqa
Note that to zeroth order, 
the new system is in state $\ket{-}\ket{00}$, as 
expected:
\beq
\Omega_{a}(dt)\ket{S^{+}} \propto \ket{S^{-}}.
\eeq
Moreover, the rate for this detection to occur is
\beq
 \bra{S^{+}} \Omega_{a}\dg(dt)\Omega_{a}(dt) \ket{S^{+}}/dt,
\eeq
which to zeroth order evaluates  to $\gamma/4$, as expected from the dressed 
atom model.

Returning to the more complete expression (\ref{morcom}) for the system state 
following an $a$ detection,  there is an error due to the second term, 
which is the amplitude for the system to jump to the wrong dressed 
state. The magnitude of this error is clearly
\beq
\epsilon_{\rm wrong} \sim \frac{\Gamma^{2}}{\Omega^{2}}.
\eeq
As well as this error, the correlation between the $b$ cavity and the atomic 
state is already present in the third term $\ket{+}\ket{01}$, with the 
same sign as in the entangled state $\ket{S^{-}}$. Thus any subsequent 
detection through the $b$ cavity will have the same effect as if the 
system were in state $\ket{S^{-}}$, namely to put the system back in 
the state $\ket{+}\ket{00}$. However, if a photon rejected by both 
cavities is detected before the third term has decayed to its 
stationary value, the 
system will not jump into the correct approximate dressed state. This 
can be easily verified from \erf{morcom}. Since 
the rate of such detections is of order $\gamma/4$ and the square of 
the amplitude for the third term decays at rate $2\Gamma$, the error 
introduced by this transient scales as
\beq
\epsilon_{\rm transient} \sim \frac{\gamma}{8\Gamma}.
\eeq

A different sort of error occurs when a photodetection occurs which 
would be forbidden by the dressed state model. 
In the present situation, when the system starts in the state 
$\ket{S^{+}}$, the forbidden detection is through cavity $b$. Since
the (unnormalized) state conditioned on this detection is
\beq
\Omega_{b}(dt)\ket{S^{+}} 
= \rt{dt\gamma/2 }\ket{+}\left( \frac{-i\Gamma}{\rt{2}\Omega}\ket{00} 
-\frac{\sqrt{\gamma}}{4\sqrt{2\Gamma}}\ket{10}\right),
\eeq
the probability for this detection to occur scales as
\beq \label{epforb}
\epsilon_{\rm forbidden} \sim \frac{\Gamma^{2}}{2\Omega^{2}} + 
\frac{\gamma}{32\Gamma}.
\eeq

Turning now to the other allowed detection, we find that to zeroth 
order
\beq
\Omega_{1}(dt)\ket{S^{+}} \propto \ket{S^{+}},
\eeq
and the rate of this process is again $\gamma/4$. Similarly, if the 
atom starts in the state $\ket{S^{-}}$ the to zeroth order
\bqa
\Omega_{b}(dt)\ket{S^{-}} &\propto& \ket{S^{+}}, \\
\Omega_{1}(dt)\ket{S^{-}} &\propto& \ket{S^{-}},
\eqa
and the rates are $\gamma/4$, while the probability of a forbidden 
detection through cavity $a$ is similar to the expression 
(\ref{epforb}). Thus for $\gamma \ll \Gamma \ll \Omega$, 
 the system almost always jumps between the states $\ket{S^{\pm}}$. 
This is confirmed by the numerical simulations shown in Fig.~4.
 
 The final 
source of error is that in these states the atomic state is not the pure 
dressed state expected. Rather, from \erf{466} and \erf{467}, 
the orthogonal dressed state is 
entangled with the cavity states, with amplitude 
$\rt{\gamma}/2\rt{\Gamma}$. Thus the probability for not finding the 
atom in the expected dressed state scales as
\beq
\epsilon_{\rm entangled} = \frac{\gamma}{4\Gamma}.
\eeq
There is another error introduced by the fact that the atomic states 
$\ket{h},\ket{l}$ differ from the dressed states $\ket{+},\ket{-}$ by 
an amount of order $(\gamma/\Omega)^{2}$. However since $\Gamma \gg 
\gamma$ this is negligible compared to other errors noted above, such 
as $\epsilon_{\rm wrong}$.

To determine the total probability for deviation from the predictions 
of the simple dressed state model, we add the four sources of error 
discussed above.  The result is
\beq
\epsilon_{\rm total} = \alpha \frac{\Gamma^{2}}{\Omega^{2}} + 
\beta \frac{\gamma}{4\Gamma}, \label{epstot}
\eeq
where $\alpha$ and $\beta$ are imprecisely known parameters of order 
unity. Minimizing the total error for fixed $\Omega$ and $\gamma$ 
implies that the filters should be chosen to have a linewidth scaling 
as
\beq
2\Gamma \sim \Omega^{2/3}\gamma^{1/3}, \label{bestlw}
\eeq
where this expression would be exact for $\alpha=\beta$. This 
optimal scaling is interesting in that it differs from the geometric mean 
$\Omega^{1/2}\gamma^{1/2}$ which is what one might have guessed.
Substituting this back into \erf{epstot} gives
\beq \label{esptot}
\epsilon_{\rm total} \sim \frac{3}{4}\left( 
\frac{\gamma}{\Omega}\right)^{2/3},
\eeq
where this expression would be exact for 
$\alpha=\beta=1$. Thus with $\Omega = 700\gamma$, which is readily 
achievable experimentally, the atomic dynamics would agree with those 
of the dressed state model with an accuracy of about $99\%$. 
%It would not agree 
%with the Teich and Mahler model at all.

Accurately testing these results via stochastic simulations
with reasonable computational resources would require 
a long time, or highly specialized
algorithms and extensive coding. However, the scaling laws can be 
tested in the following approximate way. From Fig.~4 we see that the 
state after a sideband detection alters little until the next detection into a 
different sideband. 
Therefore if we calculate the average state of the atom after a particular 
sideband detection we should get a decent estimate of how close the 
atom usually is to a dressed state. The way this average state can be 
calculated is presented in Appendix A. Let us consider a 
high-frequency sideband detection (through cavity $a$), 
and call this average state 
$\rho_{a}$. We expect this state to be close to the low-energy dressed 
state $\ket{-}\bra{-}$. Therefore we can define the approximate probability of 
error as
\beq
\epsilon_{\rm app} = \bra{+}\rho_{a}\ket{+}.
\eeq

An expression for this probability for error is derived in Appendix A, and a 
perturbation expansion 
in $\gamma/\Gamma$ and $\Gamma/\Omega$ 
yields 
\beq
\epsilon_{\rm app} \simeq 
\frac{5\Gamma^2}{4\Omega^2} + \frac{\gamma}{8\Gamma} .
\eeq
Comparing to expression (\ref{epstot}) above shows that we have 
$\alpha = 5/4$ and $\beta = 1/2$, which are of order unity as expected. 
Since the two methods for calculating the probability of error give the 
same scaling, 
we can be confident in the final results of \erf{bestlw} and 
\erf{esptot}.

\section{Two-State Jumping}

We have found that the dressed atom theory, rather than the 
TM theory, well approximates
the evolution of the atom under perfect spectral detection 
in the high-driving limit $\Omega\gg \gamma$. 
However, this does not prove that the 
TM theory does not does not describe the atomic dynamics under some 
other detection scheme which Teich and Mahler failed to identify.
To investigate this question we turn now to the second of the three 
features of the TM model listed in Sec.~\ref{secfea}. That is,
in steady state, the atomic state is always one of 
	two fixed pure states.
	
\subsection{Homodyne Detection}

Since we are seeking a physical detection scheme which will yield
two-state jumps, the theory of quantum trajectories expounded in 
Sec.~IV~A is again the appropriate theory. For the system to remain a 
pure state it is necessary for the detection scheme not to entangle 
the atomic state with any other systems (as occurs in spectral 
detection). Thus we do not require cascaded 
quantum systems theory and instead consider measurement operators 
$\Omega_{\alpha}(dt)$ in the Hilbert space of the 
atom alone. 

With this restriction it might be thought that the only 
option is then the measurement operators
\bqa
\Omega_0(dt) &=& 1 - i\frac{\Omega}{2}\sigma_{x}dt - 
\frac{\gamma}{2}\sigma\dg \sigma dt, \\
\Omega_1(dt) &=& \rt{\gamma dt}\sigma. 
\eqa
discussed in Sec.~IV~A, which correspond to direct detection and give 
rise to quantum jumps quite unlike those of the TM theory. 
However this is not the case. Although the master equation (\ref{me1}) 
is invariant under the transformation given by Eqs.~(\ref{tf1}) and (\ref{tf2}),
the measurement operators of Eqs.~(\ref{O1}) and (\ref{O0}) are not 
invariant. In the context of the two-level atom, the master equation
\beq \label{me2ag}
\dot{\rho} = -i\frac{\Omega}{2}[\sigma_{x},\rho] + \gamma {\cal D}[\sigma]\rho
\eeq
can be unraveled by any of a family of measurement operators 
parameterized by the complex number $\mu$,
\bqa
\Omega_0(dt) &=& 1 - \left(i\frac{\Omega}{2}\sigma_{x} + 
\frac{\gamma}{2}\sigma\dg \sigma + \mu^{*}\gamma\sigma + \frac 
{\gamma|\mu|^{2}}{2} \right) dt, \label{homsmooth}\\
\Omega_1(dt) &=& \rt{\gamma dt}(\sigma+\mu).
\eqa
Direct detection is recovered by setting $\mu=0$.

The transformation parameterized by $\mu$ can be physically performed 
simply by adding a coherent amplitude to the field radiated by the 
atom before detecting it. If the atom radiates into a beam as 
considered previously, this can be achieved by mixing it with a 
resonant local oscillator at a beam splitter, as shown in Fig.~5. 
This is known as homodyne detection.
The transmittance of the beam splitter must be close to unity and
the local oscillator strength chosen such that the transmitted field 
(in the absence of the atom) would have a  photon 
flux  equal to $\gamma|\mu|^{2}$. The 
phase of $\mu$ is of course defined relative to the field driving the 
atom.

Since our aim is for the atom to remain in one of two fixed pure 
states, except when it jumps, we must examine the fixed points 
(i.e.~eigenstates) of the 
operator $\Omega_{0}(dt)$. It turns out that it has two fixed states, 
such that if ${\rm Re}[\mu]\neq 0$, 
one is stable and one unstable. For ${\rm Re}[\mu]>0$, 
the stable fixed state is
\beq
\ket{\tilde{\psi}_{\rm s}^{\mu}} = \left(
\sqrt{\Omega^{2}- 2i\Omega\gamma\mu^{*}-\frac {\gamma^{2}}{4}} 
+ \frac {i\gamma}{2} \right)\ket{g} + 
\Omega\ket{e}. \label{stablefix}
\eeq
Here the tilde denotes an unnormalized state. The corresponding 
eigenvalue is
\beq
\lambda_{\rm s}^{\mu} = -\gamma\frac{1+2|\mu|^{2}}{4}-\frac i2 
\sqrt{\Omega^{2}- 2i\Omega\gamma\mu^{*}-\frac{\gamma^{2}}{4} }.
\eeq
The unstable state and eigenvalue are found by replacing the square 
root by its negative. 

\subsection{Adaptive Homodyne Detection} \label{ahd}

Let us say $\mu=\mu_{+}$, with ${\rm Re}[\mu_{+}]>0$, 
and assume the system is in the appropriate stable state 
$\ket{\psi_{\rm s}^{+}}$. When a jump occurs the new state of the 
system is proportional to
\beq
\Omega_{1}^{+}\ket{\psi_{\rm s}^{\mu}} \propto (\sigma+\mu_{+})\ket{\psi_{\rm 
s}^{+}}.
\eeq
The new state will obviously be different from $\ket{\psi_{\rm 
s}^{+}}$ and so will not remain fixed. This is in contrast to what we 
are seeking, namely a system which will remain fixed between jumps. 
However, let us imagine that immediately following the detection, the 
value of the local oscillator amplitude $\mu$ is changed to some new 
value, $\mu_{-}$. This is an example of an {\em adaptive} measurement 
scheme \cite{Wis95c,Wis96a}, in that the parameters 
defining the measurement depend upon 
the past measurement record. We want this new $\mu_{-}$ to be chosen 
such that the state $(\sigma+\mu_{+})\ket{\psi_{\rm 
s}^{+}}$ is a stable fixed point of the new $\Omega_{0}^{-}(dt)$. The 
conditions for this to be so will be examined later. If they are 
satisfied  then the 
state will remain fixed until another jump occurs. This time the new 
state will be proportional to
\beq
(\sigma + \mu_{-})(\sigma+\mu_{+})\ket{\psi_{\rm s}^{+}}
= [\mu_{-}\mu_{-} + (\mu_{-}+\mu_{+})\sigma]\ket{\psi_{\rm s}^{+}}.
\eeq
If we want jumps between just two states then we require this to be 
proportional to $\ket{\psi_{\rm s}^{+}}$. Clearly this will be so 
if and only if
\beq
\mu_{-} = -\mu_{+}. \label{mupm}
\eeq

Writing $\mu_{+}=\mu$, we now return to the condition that 
$(\sigma+\mu_{+})\ket{\psi_{\rm s}^{+}}$ be an the stable fixed state of 
$\Omega_{0}^{-}(dt)$. From \erf{stablefix}, and using \erf{mupm}, 
this gives the relation 
\beq
\sqrt{\Omega^{2}+ 2i\Omega\gamma\mu^{*}-\frac {\gamma^{2}}{4}}
= \sqrt{\Omega^{2}- 2i\Omega\gamma\mu^{*}-\frac {\gamma^{2}}{4}}
-\frac{\Omega}{\mu}.
\eeq
This has just two solutions,
\beq
\mu_{\pm} = \pm \frac 12,
\eeq
which, remarkably, are independent of the ratio $\gamma/\Omega$.
The stable and unstable fixed states for this choice are
\bqa
\ket{\psi_{\rm s}^{\pm}} &=& 
\frac{\pm\Omega-i\gamma}{\sqrt{2\Omega^{2}+\gamma^{2}}}\ket{g} - 
\frac{\Omega}{\sqrt{2\Omega^{2}+\gamma^{2}}}\ket{e}, \\
\ket{\psi_{\rm u}^{\pm}} &=& 
\frac{1}{\sqrt{2}}\ket{g} \pm \frac{1}{\sqrt{2}}\ket{e} ,
\eqa
and the corresponding eigenvalues are
\bqa
\lambda_{\rm s}^{\pm}& =& -\frac{\gamma}{8} \pm \frac{i\Omega}{2} ,\\
\lambda_{\rm u}^{\pm}& =& -\frac{5\gamma}{8} \mp \frac{i\Omega}{2}. 
\eqa

Note that the unstable states $\ket{\psi_{\rm u}^{\pm}}$ 
are simply the dressed states 
$\ket{\pm}$ defined in \erf{ds}. In the limit $\Omega \gg \gamma$, 
the stable states $\ket{\psi_{\rm s}^{\pm}}$ become equal to orthogonal 
dressed states $\ket{\mp}$. Like the states $\ket{h},\ket{l}$ defined 
in Eqs.~(\ref{keth}), (\ref{ketl}), they differ from the dressed 
states by an amount of order $(\gamma/\Omega)^{2}$. However they are 
different states, as their locus on the Bloch sphere in Fig.~6 shows.
In any case, the system evolution in steady state, jumping between 
the stable states, corresponds closely to the 
dressed state evolution.  
 It is shown in Appendix B that the system rapidly 
reaches this stationary evolution from an arbitrary initial condition.

Unlike in spectral detection, the atomic state is not entangled with 
any other system and does jump cleanly from one pure state to 
another. The total error, the probability for the atom not to be in 
the nearest dressed state, is just
\beq
\epsilon_{\rm total} = \left| \ip{+ | \psi^{+}_{\rm s}}\right|^{2} 
= \frac{\gamma^{2}}{4\Omega^{2}+2\gamma^{2}},
\eeq
which goes as $(\gamma/2\Omega)^{2}$ in the limit of strong 
driving. Note that as $\gamma/\Omega$ goes to zero, this error 
goes to zero much faster (with power $2$ compared 
to power $2/3$) than the corresponding minimum error for spectral detection 
in \erf{epstot}.

This adaptive measurement scheme described above 
would be relatively easy to implement 
experimentally (assuming that the problem of collecting all of the atomic 
fluorescence has been solved.) It requires simply an amplitude inversion
of the local oscillator after each detection. The other difference from 
usual homodyne detection is that the transmitted 
local oscillator intensity is very small: it corresponds
to half the photon flux of the atom's fluorescence if the atom were 
saturated. In either stable fixed state, 
the actual photon flux entering the detector in this 
scheme is
\beq
\ip{\psi_{\rm s}^{\pm}|(\mu_{\pm}+\sigma\dg)(\mu_{\pm}+\sigma)|\psi_{\rm 
s}^{\pm}} = \frac{\gamma}{4},
\eeq
which  is 
again independent of $\Omega/\gamma$. This rate is also of course the 
rate for the system to jump to the other stable fixed state. 
The rate $\gamma/4$ is the same as the rate of 
state-changing transitions in the dressed atom model. However, 
since there are no detections which leave the atomic state unchanged, 
the total rate of photedetections is half that of the dressed atom or 
TM model.
%However, once again, the states between which the atom jumps are 
%(approximately) the dressed states, {\em not} the diagonal states of 
%the TM model. 

\section{Orthogonal Jumping}

The final detection scheme we will examine in this paper is quite 
similar to the two-state detection scheme. Rather than reproducing the 
two-state feature of the TM model, it aims to reproduce instead the 
third feature listed in Sec.~\ref{secfea}. That is,
it will  ensure that at all times, the state after a jump is orthogonal to the 
one before. This can be achieved using homodyne detection as in the 
two-state model. The condition is 
\beq
\bra{\psi}(\sigma + \mu)\ket{\psi} = 0,
\eeq
or 
\beq \label{ojmu}
\mu(t) = - \bra{\psi(t)}{\sigma}\ket{\psi(t)}.
\eeq

Like two-state jumping, this orthogonal jumping clearly 
requires an adaptive measurement scheme 
in that the amplitude (in this case intensity and phase) of the local 
oscillator will depend on the previous measurement history, which will 
determine the current state $\ket{\psi(t)}$. But since in this case the 
amplitude $\mu$ is found directly from $\ket{\psi(t)}$, it will also 
depend on the initial state of the system at some time in the past 
$\ket{\psi(0)}$. That is, it requires the experimenter to know the 
initial state of the system. This is an unusual condition that will 
be discussed more later. 

\subsection{Dynamics}

The ``jump'' measurement  operator corresponding to the choice in 
\erf{ojmu} is
\beq
\Omega_{1}(dt) = \rt{\gamma dt}\left[\sigma-\ip{\sigma}(t)\right]
\eeq
From \erf{homsmooth}, the no-jump measurement operator is
\beq
\Omega_0(dt) = 1 - \left(i\frac{\Omega}{2}\sigma_{x} + 
\frac{\gamma}{2}\sigma\dg \sigma + \gamma\ip{\sigma}^{*}\sigma + \frac 
{\gamma|\ip{\sigma}|^{2}}{2} \right) dt. \label{ojsmooth}
\eeq
For $\Omega > \gamma$, 
this nonlinear operator has the following stable fixed states:
\beq
\rt{2}\Omega\ket{\theta_{\pm}} = \Omega\ket{g} + \left(i\gamma \pm  
\rt{\Omega^{2}-\gamma^{2}}\right)\ket{e}.
\eeq
Once again in the limit that $\Omega/\gamma$ becomes large, these 
approximate the dressed states, 
differing  from them by an error of order $(\gamma/2\Omega)^{2}$. 
However, as shown in Fig.~6, they are different both from the 
states $\ket{h},\ket{l}$ appearing in our analysis of spectral detection 
and the two stable states $\ket{\psi_{\rm s}^{\mp}}$ in the two-state 
jumps. Linearizing the nonlinear operator $\Omega_{0}(dt)$ about 
either of the fixed states gives the eigenvalues 
\beq \label{ojeig}
\lambda = -\frac{\gamma}{4} \pm \sqrt{ 
\left(\frac{\gamma}{4}\right)^{2} + (\gamma^{2}-\Omega^{2})}.
\eeq
For $\Omega > \rt{17/16}\gamma$, the eigenvalues are complex, 
so that the two fixed points on the Bloch 
sphere are foci of the no-jump dynamics. For $\Omega \gg \gamma$, the real 
and imaginary parts of the eigenvalues approach $-\gamma/4$ and 
$\Omega$ respectively.

Because of the nonlinearity of the evolution described by the above
$\Omega_{1}(dt)$ and $\Omega_{0}(dt)$, an analytical approach is more 
difficult in this case. Indeed, for $\Omega \sim \gamma$ we find 
numerically that the evolution is very complicated. However, for 
$\Omega \gg \gamma$, the two stable fixed states approach orthogonality. 
This means that if the system is in a stable fixed state, then 
a jump (which by construction takes the system to an orthogonal state) 
will take it close to the other stable fixed state. Thus we expect 
the two-state jumping of the previous section to be approximately 
reproduced in the strong-driving limit. This expectation is confirmed 
by the numerical simulations shown in Fig.~7. As expected from 
\erf{ojeig}, the 
state after a jump spirals towards the closest fixed state.

Once again, attempting to replicate a feature of the TM model has in 
fact replicated approximately the behaviour of the simple dressed 
atom model. The deviation of the orthogonal jump behaviour from that 
simple model can be roughly estimated as follows. First, as noted 
above, the stable 
fixed points differ from the dressed states by an error of 
order $(\gamma/2\Omega)^{2}$. Second, when a jump from 
$\ket{\theta_{+}}$ occurs the new state is such that the dressed 
state $\ket{-}$ lies midway between it and $\ket{\theta_{-}}$. The 
error immediately after such a jump is thus also $(\gamma/2\Omega)^{2}$. 
Hence we can estimate that the overall error (the probability of the 
atom not being in the expected dressed state at steady state) is 
\beq
\epsilon \sim \left(\frac{\gamma}{2\Omega}\right)^{2}.
\eeq
From numerical simulations shown in Fig.~8 we find
\beq
\epsilon \approx 3 \left(\frac{\gamma}{2\Omega}\right)^{2},
\eeq
so that the error is roughly three times greater 
than in the in the two-jump case.

\subsection{Nonlinearity and Consistency}

The ``orthogonal jump'' evolution presented here has been considered 
before. Breslin {\em et al.} \cite{BreMilWis95} 
used it to examine questions of 
information production and quantum chaos. Earlier, it was suggested by 
Diosi \cite{Dio94} as a unique way of unraveling a master equation. 

The orthogonal jump evolution differs from the other unravelings 
considered here in that it is nonlinear in the sense that the 
measurement operators $\Omega_{1}(dt), \Omega_{0}(dt)$ depend upon the 
state of the system. This is only possible if we assume that the 
experimenter who is monitoring the system knows what its initial state 
is. Another experimenter, arriving in the middle of the monitoring 
and having not communicated with the first experimenter, would at 
assign a different (mixed) state to the system. That second 
experimenter would then disagree with the first experimenter's control 
of the measurement apparatus (the local oscillator amplitude) because 
the two experimenters would assign different values of $\ip{c}$ to the 
system.

This disagreement between two observers on how the system 
will (or rather, should) evolve applies also the TM model. This was 
pointed out in Ref.~\cite{WisMil93c}, where it was argued that this 
was a fatal flaw in the consistency of the TM model. 
In the present context we can now 
see that the argument in Ref.~\cite{WisMil93c} is not wholly convincing. 
It is possible for 
different observers to disagree on the state of the system, 
but as long as one observer ``holds the reigns'' of the equipment,
the future evolution is unambiguous. 

Thus,  while the TM model fails on 
other grounds (in that its cannot be physically realized), its 
internal consistency is technically no worse than that of the orthogonal jump 
model. However, Teich and Mahler originally claimed that their model 
is ``the stochastic process which governs the time evolution of an 
individual quantum system''. That is, it is apparently supposed to represent an 
observer-independent reality. In this spirit, the disagreement of two 
observers on the dynamics of the system is still a serious problem.

\section{Conclusion}

In this work we have compared the evolution of a strongly driven, damped two 
level atom under a variety of stochastic evolution models. The 
model which served as the starting point for our investigations was 
the one due to Teich and Mahler. In the steady state, this predicts 
that the atom jumps between two orthogonal states,
\bqa
\rt{2}\ket{\phi_{1}} &=& \ket{g}-i\ket{e} \label{phi1a},\\
\rt{2}\ket{\phi_{2}} &=& \ket{g}+i\ket{e} \label{phi2a},
\eqa
with rates approximately equal to $\gamma/4$. 
Here $\gamma$ is the spontaneous emission rate for the atom. The 
states $\ket{\phi_{1}},\ket{\phi_{2}}$ are 
the states which diagonalize the stationary state matrix for the 
system in the limit of strong driving.

The TM model is supposed to be an objective description of the 
behaviour of individual quantum systems. However, this objectivity 
runs counter to one of the fundamental features of quantum mechanics, 
entanglement. A fluorescent atom becomes entangled with the state of 
the field into which it emits, and different ways of monitoring that 
field will give different information about the atom and hence 
collapse the atom into different states. These different 
processes are called unravelings of the master equation of the atom. 
The question we then posed is, can any unraveling reproduce the theory 
of Teich and Mahler?

The short answer is no. The long answer is far more interesting. 
Individual features of the TM model can be mimicked. First, Teich and 
Mahler claimed that the jumps of their atom corresponded to the 
emission of photons with different frequencies. We modeled this 
process using the dressed atom model (Sec.~III) and then a more exact and far 
more complicated method explicitly including the filters used for 
distinguishing the different frequencies (Sec.~IV). The rate and ordering 
of jumps were roughly as predicted by the TM model. Then (Sec.~V), we 
derived the measurement scheme which ensures that the atom, in steady 
state, jumps between precisely two states (as in the TM model).
Last (Sec.~VI), we derived the scheme which ensures that when the atom jumps to 
a new state, it is orthogonal to the old one (as in the TM model).

What is interesting is that in all of these cases the atom spends most 
of its time close to one of the following two states
\beq \label{dsa}
\rt{2}\ket{\pm} = \ket{g}\pm \ket{e}.
\eeq
These we have called, in a slight abuse of terminology, the dressed 
states of the atom. That is because in the simple dressed-state 
theory, these are precisely the states the atom jumps between for a 
coherent driving field. The probability for error, that is the 
probability for the system to be in a state other than the expected 
one of these dressed states, depends on the monitoring scheme. 
For the three schemes 
considered here, the error probabilities are respectively
\bqa
\epsilon_{\rm spectral} &\sim& 
\frac{3}{4}\left(\frac{\gamma}{\Omega}\right)^{2/3} \\
\epsilon_{\rm two-state} &\simeq& 
\frac{1}{4}\left(\frac{\gamma}{\Omega}\right)^{2}, \\
\epsilon_{\rm orthogonal} &\approx& \frac{3}{4} 
\left(\frac{\gamma}{\Omega}\right)^{2},
\eqa
where $\Omega \gg \gamma$ is the Rabi frequency of the driving field.

These results are in flat 
contradiction to those of the TM model. The dressed states are as far 
away as possible from the diagonal states $\ket{\phi_{1}},\ket{\phi_{2}}$.
But our purpose here is not to bludgeon the TM model by repeated 
instances of nonphysicality, but to marvel at the fact that whenever 
one tries to mimic it, one ends up instead mimicking the evolution of 
the dressed state model. It seems as if jumping between the dressed 
states is the evolution which the atom ``wants to do''. Of course one 
can force it to behave otherwise. Direct photodetection will result 
in quite different evolution. But attempting to make the evolution 
simple seems to lead inevitably to the dressed state jumping.

The lesson here is that diagonal states, that is, the states which 
diagonalize the state matrix, have no relation to the states the 
system prefers. In the case of a resonantly driven two-level atom 
the preferred states seem to be the dressed states. Lest the reader be 
annoyed at our repeated attribution of state preference to a quantum system, 
we point out that this terminology is widely used in the study of 
decoherence and the classical limit \cite{Zur93}. A more technical 
term is ``environmentally-induced superselection'', which emphasizes 
that the preference of certain quantum states is a property of the 
environment of the system, as well as the system itself.

A two-level atom is such a small quantum system that it might be thought 
that there is no point considering a classical limit. This is a fair 
comment if one supposes that a classical limit must be a deterministic 
limit. But if one is prepared to accept a stochastic classical model, 
than it seems that something like the
dressed state model is the appropriate limit 
for a strongly driven atom. 

Of the measurement schemes analyzed here, 
the two-state scheme of Sec.~V gives the best approximation 
to the evolution arising from the dressed state model. 
A rigorous way of defining the preferred 
unraveling of a master equation has been formulated recently by one 
of us \cite{WisVac98}, using the concept of robustness. Preliminary 
work suggests that the two-state jumping scheme is in fact the 
most robust unravelling of the resonance fluorescence master equation.
This will be pursued in a future work. 

\acknowledgments
We would like to acknowledge support by the Australian Research 
Council.

\appendix
\section{Average State Conditioned on a Filtered Photon Detection}

In this appendix we show how to calculate the average state of the 
atom given that a  detection through a particular filter has just 
occurred. This calculation requires consideration of only a single 
filter cavity, which is described by the master equation
\bqa
\dot{W} &=& -i\left[ ({\Omega}/{2})\sigma_{x} ,W\right] + 
\gamma {\cal D}[\sigma]W \nn \\
&& -\, i\left[ \omega_{a}a\dg a,W\right] + 2\Gamma {\cal D}[a] W \nn \\
&& +\, \sqrt{\Gamma\gamma} \left( a W \sigma\dg + \sigma W a\dg - a\dg \sigma
W - W a \sigma \dg  \right) ,\label{one}
\eqa
which is just a rearrangement of \erf{casc1}. 
The first line is the atomic dynamics, the second line is
the cavity's free dynamics and the third line is the coupling of the 
atom to the cavity.

The average state of the atom immediately after the detection of a 
photon passed by the cavity is  
\beq
{\rho}_{a}  = \frac{{\rm Tr}_{\rm cav} [ a W a \dg]}{{\rm Tr} [ a W a 
\dg]},
\eeq
where here $W$ is the stationary solution of \erf{one}. To determine 
this state, we first  trace out the field entirely to obtain the familiar master
equation for the TLA of,
\begin{equation}
\frac{d \rho_{0}}{dt} = -i\left[ ({\Omega}/{2})\sigma_{x},\rho_{0} \right] 
+ \gamma {\cal D}[\sigma]\rho_{0} \equiv {\cal L} \rho_{0}.
\end{equation} 
Writing the atomic state matrix in  the Bloch representation,
\begin{equation}
\rho_{0} = \frac{1}{2} \left( p_{0} I + x_{0} \sigma_x + y_{0} \sigma_y + z_{0} \sigma_z 
\right),
\end{equation}
 the relaxation superoperator takes the form
\begin{equation}
{\cal L} = 
\left(
\begin{array}{cccc}
 0 & 0 & 0 & 0  \\ 
 0 & -\frac{1}{2}\gamma & 0 & 0 \\
 0 & 0 & -\frac{1}{2}\gamma & -\Omega \\
 -\gamma & 0 & \Omega & -\gamma 
\end{array}
\right),
\end{equation}
acting on the vector $(p_{0},x_{0},y_{0},z_{0})^{T}$. 
Assuming that the state is 
initially normalized ($p_{0}=1$), the stationary solution is
\begin{equation}
(p_0,x_{0},y_{0},z_{0}) = \left(1,0,
\frac{2 \Omega\gamma}{\gamma^{2} + 2 \Omega^2},\frac{- \gamma^{2}} { \gamma^{2} 
+ 2 \Omega^2} \right) .
\end{equation}

Next, we calculate the steady-state value of $\rho_1 = {\rm Tr}_{\rm
cav} [a W]$. From \erf{one} we get
\begin{equation}
\frac{d \rho_1}{dt} = {\cal L} \rho_1 -i \omega_a \rho_1 - \Gamma \rho_1 -
\sqrt{\Gamma\gamma}\, \sigma \rho_{0}.
\end{equation}
Converting the operator $\sigma \rho_{0}$ into the 
Bloch vector representation, 
we find that the steady-state representation of $\rho_1$ is
\begin{eqnarray}
\rho_1 &=& \frac{\sqrt{\Gamma\gamma}}{2} \left( {\cal L} - i \omega_a - \Gamma
\right)^{-1} 
\nl{\times} \left( x_0- i y_0 , p_0+z_0 , -i
p_0 -i z_0 , - x_0+iy_0 \right)^T 
\end{eqnarray}

We can now repeat this process for
$\rho_2 = {\rm Tr}_{\rm cav} [ a W a\dg ]$.  Tracing over Eq.~\ref{one}
gives
\begin{equation}
\frac{ d \rho_2}{dt} = {\cal L} \rho_2 - 2 \Gamma \rho_2 - \sqrt{\Gamma\gamma} \left(
\rho_1 \sigma \dg + \sigma \rho_1\dg \right).
\end{equation}
Thus the value of $\rho_2$ in terms of $\rho_1$ is given by
\begin{equation}
\rho_2 = \frac {\sqrt{\Gamma\gamma}}{ {\cal L} - 2 \Gamma } \left( x_1^r  -y_1^i,
p_1^r + z_1^r, -p_1^i -z_1^i, -x_1^r + y_1^i  \right)^T,
\end{equation}
where the $r$ and $i$ superscripts indicate real and imaginary parts.
Substituting in the above expression for $\rho_{1}$ and normalizing 
thus gives $\rho_{a}$ as required. 

Due to the algebraic complexity of the exact matrix inverses, it is
more instructive to examine the perturbative solution in the limit 
$\gamma/\Gamma,\Gamma/\Omega \ll 1$. 
We are particularly interested then in the case where the cavity is 
tuned to a sideband, with $\omega_{a} = \Omega$.
Writing $\rho_{2}$ in the Bloch 
representation, the  
probability of error (which is the quantity we ultimately require), 
can be written as
\beq
\epsilon_{\rm app} = \frac{\bra{+}\rho_{2}\ket{+}}{{\rm 
Tr}[\rho_{2}]} = \frac{x_{2}+p_{2}}{2p_{2}}.
\eeq
Then, using the symbolic series expansion capability of MATHEMATICA, 
we find
\beq
\epsilon_{\rm app} = \frac{\gamma}{8\Gamma} + 
\frac{5\Gamma^{2}}{4\Omega^{2}} - \frac{\gamma^{2}}{32\Gamma^{2}}
+ \frac{7\Gamma\gamma}{8\Omega^{2}} + \ldots.
\eeq
The two leading terms in this expansion are quoted in the main text.

Not relevant to the problem at hand, but nevertheless interesting, is 
the question of what would happen in a different limit, namely 
$\Gamma \ll \gamma \ll \Omega$. Here the cavity is so narrow that any 
photon it passes is definitely from one peak of the Mollow triplet 
only (assuming the cavity is tuned appropriately). It might be thought 
that this would condition the state of the atom very well In fact 
the perturbative result obtained for $\omega_a = \Omega + \Delta$, 
with $\Delta \ll Omega$, is
\begin{equation} \label{rho2}
	\rho_{2} \propto  \frac{
	\left(\smallfrac{3}{4}\gamma\right)^{2}}{\Delta^2 +
\left(\smallfrac{3}{4}\gamma\right)^2} \left[ 1 -  
\frac{4\Gamma}{\gamma} \sigma_{x} \right].
\end{equation}
The Lorentzian lineshape here is as expected from the analysis of 
Mollow \cite{Mol69}, but the linear 
dependence of $x_{2}$ upon $\Gamma/\gamma$ shows that the narrower the cavity 
the worse the conditioning of the state. This can be understood as 
follows.  
The detection of a photon having passed through the cavity 
actually gives information 
about the atom at the time the photon entered the cavity. The transit 
time of the photon through the cavity has an 
exponential distribution $2\Gamma e^{-2\Gamma 
t}$.  If the atom did jump into the negative $\sigma_{x}$ eigenstate at time 
$0$, then by time $t$ the expectation value of $\sigma_{x}$ would 
be equal to $-e^{-\gamma t /2}$. The average value of $\sigma_{x}$ 
at the time of detection would thus be equal to
\beq
x_{a} = -\int_{0}^{\infty} 2\Gamma e^{-2\Gamma t - \gamma t/2} dt  \simeq 
-\frac{4\Gamma}{\gamma}
\eeq
for $\Gamma \ll \gamma$. This agrees with the result in \erf{rho2} 
once it is normalized.

\section{Transient Behaviour of the Two-Jump Model}

From Sec.~\ref{ahd} it is evident that an atom prepared in one of 
the stable states 
$\ket{\psi_{\rm s}^{\pm}}$ will always jump between the 
two stable states. However it remains to be shown that the atom will 
relax towards this behaviour for any initial state.
This can be done as follows. Consider the local oscillator set 
to $\mu_{+}$ and the atom in a state $\ket{\psi(0)} = p 
\ket{\psi^+_{\rm s}} + 
q \ket{\psi^+_{\rm u}}$.  If the system then evolves for a time $t$ before 
making a jump, its unnormalized state $\ket{\tilde\psi(t)}$ after the 
jump is
\bqa
\ket{\tilde \psi(t)} &=& \left[ \frac{p \exp(\lambda^+_{\rm s} t) }{2} + 
\frac{2q \exp (\lambda^+_{\rm u} t)}{2\Omega + i\gamma}\right] \ket{\psi^-_{\rm 
s}}\nn \\
&&+ \, \frac{(2\Omega-i\gamma)q \exp (\lambda^+_{\rm u} t)}{2(2 \Omega + 
i\gamma)} \ket{\psi^-_{\rm u}}.
\label{3.14}
\eqa
Here $\ip{\tilde \psi(t) | \tilde \psi(t)}$ equals the probability 
per unit time for the jump to occur. 

Two-state jumping will be stable if the average jump, at time $t$, causes 
$|\langle \psi(t) | \psi^-_{\rm u} \rangle |^2$ to decrease.
Conveniently there is no need to examine the case for $-\mu$ because 
of the symmetry of the situation.  The quantity we wish to calculate 
is the average of $|\langle \psi(t) | \psi^-_{\rm u} \rangle 
|^2$, weighted by the probability of the jump occurring at time 
$t$. The result is
\begin{eqnarray}
{\rm E}\left[|\langle \psi(t) | \psi^-_{\rm u} \rangle |^2 \right] &=& 
\int_0^\infty |\langle \psi(t) | \psi^-_{\rm u} \rangle |^2 \langle \tilde 
\psi(t) | \tilde \psi(t) \rangle dt  \\
&=& \int_0^\infty |\langle \tilde \psi(t) | \psi^-_{\rm u} \rangle |^2 
dt.
\eqa
For $\Omega \gg \gamma$, the stable and unstable states are nearly 
orthogonal so we have 
\beq
{\rm E}\left[|\langle \psi(t) | \psi^-_{\rm u} \rangle |^2 \right] 
\simeq \int_{0}^{\infty} \frac{|q\exp(\lambda^+_{\rm u} t)|^2 dt}{4} %\nn \\
= \frac{|q|^2}{5}
\label{3.15}
\eeq
On average, then, after each jump the probability for the atom to be 
in the unstable state is reduced to a fifth of its size prior to the jump. 
Thus we can conclude that the two-state evolution is stable.

\newpage
\begin{figure}
	\epsfig{figure=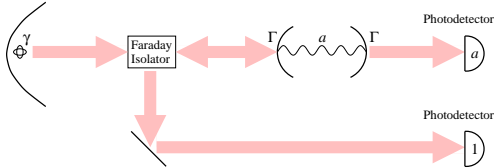,width=65mm}
\caption{\narrowtext Diagram of the experimental configuration 
	for obtaining the fluorescence of an atom as a beam, and passing it 
	through a filter cavity.}
	\protect\label{fig1}
\end{figure}

\begin{figure}
	\epsfig{figure=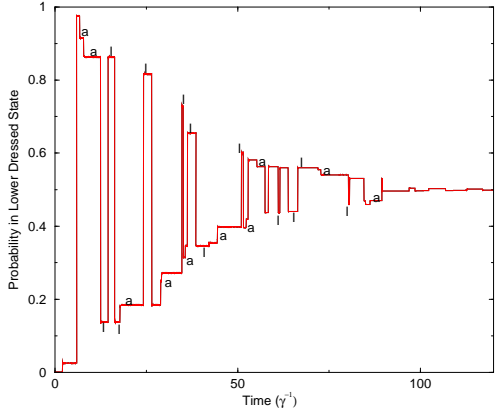,height=65mm,angle=-90}
	\caption{\narrowtext Probability for the atom to be in the dressed 
	state $\ket{-}$ for a typical observation record from 
	spectral detection using a single cavity tuned 
	to the resonant frequency of the atom. The symbol {\sf a} denotes the 
	state following the detection of a photon at photodetector $a$ 
	and {\sf l} at photodetector $1$ (see Fig.~1).
	The parameters used are 
	$\Omega/\gamma = 50$, $\Gamma/\gamma = 8$.}
	\protect\label{fig2}
\end{figure}

\begin{figure}
	\epsfig{figure=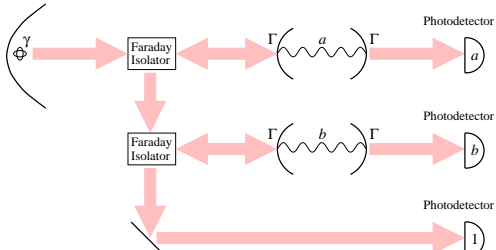,width=65mm}
	\caption{\narrowtext Diagram of the experimental configuration 
	for passing the fluorescence of an atom through two filter cavities. }
	\protect\label{fig3}
\end{figure}

\begin{figure}
	\epsfig{figure=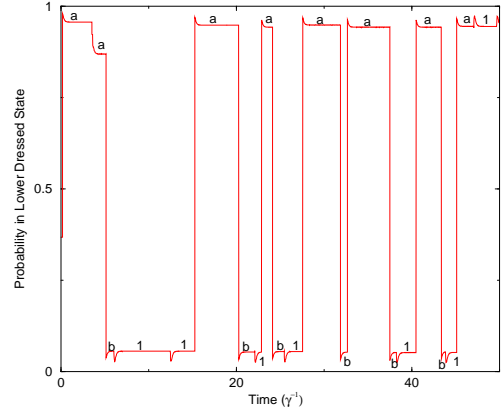,height=65mm,angle=-90}
	\caption{\narrowtext Probability for the atom to be in the dressed 
	state $\ket{-}$ for spectral detection using two cavities tuned 
	to the sidebands of the Mollow triplet, $\omega_{0}\pm \Omega$. 
	The symbol {\sf a} denotes the 
	state following the detection of a photon at photodetector $a$, 
	{\sf b} at photodetector $b$, and {\sf l} 
	at photodetector $1$  (see Fig.~3). 
	The parameters used are 
	$\Omega/\gamma = 50$, $\Gamma/\gamma = 8$.}
	\protect\label{fig4}
\end{figure}

\begin{figure}
	\epsfig{figure=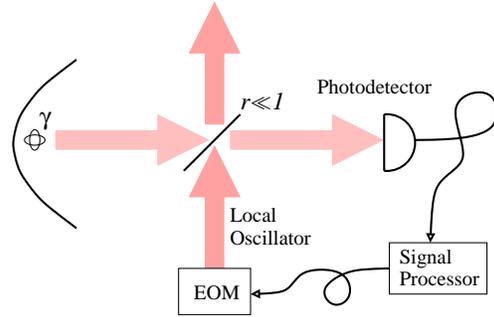,width=65mm}
	\caption{\narrowtext Diagram of the experimental configuration 
	for a homodyne measurement of the fluorescence of an atom. The 
	amplitude of the local oscillator is assumed to be variable as a 
	function of time, determined by an electro-optic modulator (EOM).
	The modulator can be controlled by the experimenter using the 
	results of the measurement.}
	\protect\label{fig5}
\end{figure}

\begin{figure}
	\epsfig{figure=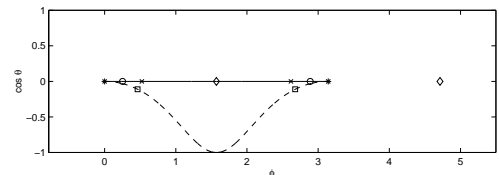,width=65mm}
	\caption{\narrowtext Locus of various states of the Bloch sphere: 
	$\ket{h},\ket{l}$ from 
	Sec.~IV for $\Omega/\gamma \in [1/2,\infty)$ (solid line and 
	circle); $\ket{\psi_{\rm s}^{\pm}}$ from Sec.~V for 
	$\Omega/\gamma \in [0,\infty)$ (dashed line and 
	square); $\ket{\theta_{\pm}}$ from Sec.~VI for $\Omega/\gamma \in [1,\infty)$ 
	(solid line and cross). 
	The circle, square and cross show the states for $\Omega = 2\gamma$. 
	Also shown are the dressed states $\ket{\pm}$ (asterisks) 
	and the high-driving 
	limit of the TM states $\ket{\phi_{1}},\ket{\phi_{2}}$ (diamonds).
	An equal area projection of the sphere in terms of the Euler angles 
	$\phi,\theta$ is used.
	}
	\protect\label{fig6}
\end{figure}

\begin{figure}
    \epsfig{figure=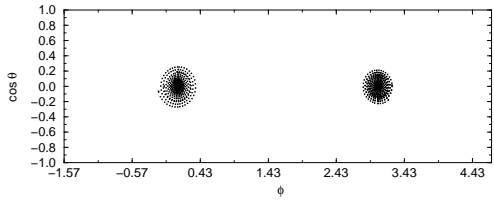,height=65mm,angle=-90}
	\caption{\narrowtext Stationary probability distribution
	of the atomic state on the Bloch sphere under  
	the orthogonal jump detection scheme. The distribution is 
	approximated by 1000 points, which are separated in time by 0.015 
	$\gamma^{-1}$. The ratio of driving to damping 
	is $\Omega/\gamma = 10$. The same equal area projection of the sphere 
	as in Fig.~6 is used, but note that the $\phi$ axis is cut at a 
	different point.}
	\protect\label{fig7}
\end{figure}

\begin{figure}
	\epsfig{figure=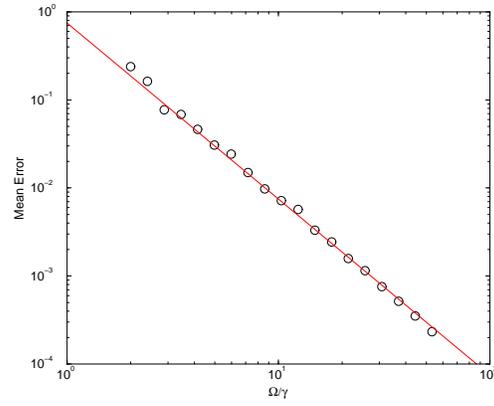,height=65mm,angle=-90}
	\caption{\narrowtext 
	Mean error (ie.~probability for the system not to be in the 
	predicted dressed state) versus
        $\Omega/\gamma$ for orthogonal jump evolution.
The fitted curve  is $\epsilon_{\rm app} = 2.99 \times ({\gamma}/{2
\Omega})^2$ }
	\protect\label{fig8}
\end{figure}

\end{multicols}

\end{document}